\documentclass[twocolumn,usenames,dvipsnames]{aastex63}
\pdfoutput=1 
\usepackage{amsmath,amstext}
\usepackage[T1]{fontenc}
\usepackage{apjfonts} 
\usepackage[figure,figure*]{hypcap}
\usepackage{lineno}
\usepackage{verbatim}
\usepackage{fancyhdr}

\hypersetup{linkcolor=red,citecolor=Blue,filecolor=cyan,urlcolor=magenta}

\newcommand{\RN}[1]{\uppercase\expandafter{\romannumeral#1}}

\usepackage{xspace}
\newcommand{\code}[1]{\texttt{#1}\xspace}
\newcommand{\targ}{COOL J1241+2219~}

\newcommand{\cref}[1]{{{#1}}}

\graphicspath{{./}{figures/}}

\shorttitle{The Brightest Galaxy in the z > 5 Universe}
\shortauthors{Khullar et al.}

\begin{document}

\title{COOL-LAMPS \RN{1}. An Extraordinarily Bright Lensed Galaxy at Redshift 5.04 \footnote{This paper includes data gathered with the 6.5m Magellan Telescopes located at Las Campanas Observatory, Chile.}}

\author[0000-0002-3475-7648]{Gourav Khullar}
\affiliation{Department of Astronomy and Astrophysics, University of
Chicago, 5640 South Ellis Avenue, Chicago, IL 60637, USA}
\affiliation{Kavli Institute for Cosmological Physics, University of
Chicago, 5640 South Ellis Avenue, Chicago, IL 60637, USA}
\author[0000-0003-2294-4187]{Katya Gozman}
\affiliation{Department of Astronomy and Astrophysics, University of
Chicago, 5640 South Ellis Avenue, Chicago, IL 60637, USA}
\affiliation{Department of Astronomy, University of Michigan, 1085 South University Drive, Ann Arbor, MI 48109, USA}
\author[0000-0003-1266-3445]{Jason J. Lin}
\affiliation{Department of Astronomy and Astrophysics, University of
Chicago, 5640 South Ellis Avenue, Chicago, IL 60637, USA}
\author[0000-0002-8397-8412]{Michael N. Martinez}
\affiliation{Department of Astronomy and Astrophysics, University of
Chicago, 5640 South Ellis Avenue, Chicago, IL 60637, USA}
\author[0000-0001-9225-972X]{Owen S. Matthews Acu\~{n}a}
\affiliation{Department of Astronomy and Astrophysics, University of
Chicago, 5640 South Ellis Avenue, Chicago, IL 60637, USA}
\author{Elisabeth Medina}
\affiliation{Department of Astronomy and Astrophysics, University of
Chicago, 5640 South Ellis Avenue, Chicago, IL 60637, USA}
\author[0000-0001-5931-5056]{Kaiya Merz}
\affiliation{Department of Astronomy and Astrophysics, University of
Chicago, 5640 South Ellis Avenue, Chicago, IL 60637, USA}
\author[0000-0002-9142-6378]{Jorge A. Sanchez}
\affiliation{Department of Astronomy and Astrophysics, University of
Chicago, 5640 South Ellis Avenue, Chicago, IL 60637, USA}
\author[0000-0002-2358-928X]{Emily E. Sisco}
\affiliation{Department of Astronomy and Astrophysics, University of
Chicago, 5640 South Ellis Avenue, Chicago, IL 60637, USA}
\author[0000-0001-8008-7270]{Daniel J. Kavin Stein}
\affiliation{Department of Astronomy and Astrophysics, University of
Chicago, 5640 South Ellis Avenue, Chicago, IL 60637, USA}
\author[0000-0002-1106-4881]{Ezra O. Sukay}
\affiliation{Department of Astronomy and Astrophysics, University of
Chicago, 5640 South Ellis Avenue, Chicago, IL 60637, USA}
\author[0000-0001-6584-6144]{Kiyan Tavangar}
\affiliation{Department of Astronomy and Astrophysics, University of
Chicago, 5640 South Ellis Avenue, Chicago, IL 60637, USA}
\author[0000-0003-1074-4807]{Matthew B. Bayliss}
\affiliation{Department of Physics, University of Cincinnati, Cincinnati, OH 45221, USA}
\author[0000-0001-7665-5079]{Lindsey E. Bleem}
\affiliation{Kavli Institute for Cosmological Physics, University of
Chicago, 5640 South Ellis Avenue, Chicago, IL 60637, USA}
\affiliation{Argonne National Laboratory, High-Energy Physics Division,
9700 S. Cass Avenue, Argonne, IL 60439, USA}
\author{Sasha Brownsberger}
\affiliation{Department of Physics, Harvard University, 17 Oxford Street, Cambridge, MA 02138, USA}
\author[0000-0003-2200-5606]{H{\AA}kon Dahle}
\affiliation{Institute of Theoretical Astrophysics, University of Oslo, P.O. Box 1029, Blindern, NO-0315 Oslo, Norway}
\author[0000-0001-5097-6755]{Michael K. Florian}
\affiliation{Observational Cosmology Lab Code 665, NASA Goddard Space Flight Center, Greenbelt, MD 20771, USA}
\author[0000-0003-1370-5010]{Michael D. Gladders}
\affiliation{Department of Astronomy and Astrophysics, University of
Chicago, 5640 South Ellis Avenue, Chicago, IL 60637, USA}
\affiliation{Kavli Institute for Cosmological Physics, University of
Chicago, 5640 South Ellis Avenue, Chicago, IL 60637, USA}
\author[0000-0003-3266-2001]{Guillaume Mahler}
\affiliation{Department of Astronomy, University of Michigan, 1085 South University Drive, Ann Arbor, MI 48109, USA}
\author[0000-0002-7627-6551]{Jane R. Rigby}
\affiliation{Observational Cosmology Lab Code 665, NASA Goddard Space Flight Center, Greenbelt, MD 20771, USA}
\author[0000-0002-7559-0864]{Keren Sharon}
\affiliation{Department of Astronomy, University of Michigan, 1085 South University Drive, Ann Arbor, MI 48109, USA}
\author[0000-0002-2718-9996]{Antony A. Stark}
\affiliation{Center for Astrophysics | Harvard \& Smithsonian, 60 Garden St, Cambridge, MA 02138}

\email{Author for correspondence: gkhullar@uchicago.edu}

\begin{abstract}
We report the discovery of COOL J1241+2219, a strongly-lensed galaxy at redshift $z$=5.043$\pm$0.002 with observed magnitude $z_{AB}=20.47$, lensed by a moderate-mass galaxy cluster at $z$=1.001$\pm$0.001. \targ is the brightest lensed galaxy currently known at optical and near-infrared wavelengths at $z$ $\gtrsim$ 5; it is $\sim$5 times brighter than the prior record-holder lensed galaxy, and several magnitudes brighter than the brightest unlensed galaxies known at these redshifts. It was discovered as part of COOL-LAMPS, a collaboration initiated to find strongly lensed systems in recent public optical imaging data. We characterise the lensed galaxy, as well as the central galaxy of the lensing cluster using ground-based $griz$JH imaging and optical spectroscopy. We report model-based magnitudes, and derive stellar masses, dust content, metallicity and star-formation rates via stellar-population synthesis modeling. Our lens mass modeling, based on ground-based imaging, implies a median source magnification of $\sim$30, which puts the stellar mass and star formation rate (in the youngest age bin, closest to the epoch of observation) at logM$_{*}$ = $10.11^{+0.21}_{-0.26}$ and SFR = $27^{+13}_{-9}$ M$_{\odot}$/yr, respectively. We constrain a star formation history for \targ consistent with constant star formation across $\sim$1 Gyr of cosmic time, and that places this galaxy on the high-mass end of the star-forming main sequence. \targ is 2-4 times more luminous than a galaxy with the characteristic UV luminosity at these redshifts. The UV continuum slope $\beta$= -2.2$\pm$0.2 places this galaxy on the blue side of the observed distribution of galaxies at $z$=5, although the lack of Ly$\alpha$ emission indicates dust sufficient to suppress this emission.

\end{abstract}


\section{Introduction} \label{sec:intro}

Strong gravitational lensing can transform typically small, faint, marginally-resolved field galaxies into highly-magnified arcs that can appear very bright. The brightest and most magnified lensed galaxies enable exquisite studies of internal galaxy morphologies down to scales of only tens of parsecs even in the most distant galaxies (e.g.,\citealp{ Bayliss_2014,Livermore_2015,Johnson_2017,Johnson_2017_2,Cornachione_2018,Rivera-Thorsen_2019,Ivison_2020,Florian_2020}). Such lensed sources are exceedingly rare, and are generally found by searching many thousands of square degrees of ground-based survey imaging (e.g.,\citealp{Allam_2007, Belokurov_2007,Koester_2010}) or by dedicated imaging programs targeting massive galaxy clusters not previously explored (\citealp{Bleem_2015,Bleem_2020,Coe_2019}). Conversely, deep observations of known strong lenses with large lensing cross-sections (c.f. the Hubble Frontier Fields; \citealp{Lotz_2017}) discover many more lensed galaxies, but these are magnitudes fainter than the rare systems found in wide searches.

\cref{Studies of the brightest lensed galaxies are complementary to studies of faint lensed galaxies, and unlensed field galaxies, offering both enhanced spatial resolution in the source plane, and sufficient photons to allow exceptional spectral resolution at useful signal-to-noise ratios.} Hence, the brightest strongly-lensed galaxies are the observational signposts with which we navigate the more abundant but much fainter deep-field samples. To date, searches of the Sloan Digital Sky Survey (SDSS; \citealp{Ahumada_2020}) have identified hundreds of bright lensed sources at $z$<3, the brightest of which typically have integrated g$_{AB}$ magnitudes of $\sim$20 \citep[e.g.,][]{Kubo_2010,Bayliss_2011,Stark_2013}.

This success is due to the depth and filter selection of the SDSS, sufficient to find the bright blue arcs that are the hallmark of UV-bright lensed galaxies at these redshifts. Many of these lensed targets have  been followed up with HST imaging that has revealed details of star-formation and structure on spatial scales down to tens of parsecs (e.g., \citealp{Johnson_2017,Cornachione_2018}). The brightest  of these lensed galaxies — a few of which predate the discoveries from the SDSS — have been followed up with high-quality spectroscopy (e.g., \citealp{Pettini_2000,James_2014,Rigby_2018a,Rigby_2018b}) that exceeds in both spectral resolution and signal-to-noise what is available for stacks of dozens or even hundreds of field galaxies (\citealp{Shapley_2003, Steidel_2016}).

Beyond $z\sim$3 however, bright lensed samples rapidly decline, and by $z\sim$5, there are only a few known strongly-lensed galaxies of even modest brightness at optical/infrared (OIR)  wavelengths; the SDSS, which has been so effective for discovery at $z\sim$2-3 has proven mostly insufficient in depths at $z\sim$5, particularly in key redder filters as the galaxies drop out of the bluer filters due to intervening IGM absorption. The brightest three $z\sim$5 spectroscopically-confirmed lensed galaxies published to date have AB magnitudes of  $i$=22.2  ($z$=4.88,  \citealp{Gladders_2002}), $i$=22.6 ($z$=4.92, \citealp{Franx_1997,Soifer_1998}), and $i$=23.3 ($z$=4.87, \citealp{Frye_2002}). The brightest was found through a wide-area search for arcs; the other two were found by targeting massive galaxy clusters. In addition, the Reionizing Lensing Cluster Survey (RELICS) program reports 5 candidate lensed galaxies with $z_{phot}$=5.5–6 and  F160W<23.6 (\citealp{Salmon_2020}), that as yet lack published spectroscopic redshifts. For comparison, the brightest galaxies at $z\sim5$ in the Hubble Ultra Deep Field (HUDF) have magnitudes fainter than $i_{AB}=25$ (\citealp{Inami_2017}).

Strong-lensing searches in the latest generation of imaging surveys, particularly the Dark Energy Camera Legacy Survey (DECaLS; \citealp{Huang_2020}), the Dark Energy Survey (DES; \citealp{Diehl_2017,Jacobs_2019}), the Kilo-Degree Survey (KiDS; \citealp{Petrillo_2019}), the Hyper Suprime-Cam Subaru Strategic Program (HSC-SSP) Survey (\citealp{Jaelani_2020}), should significantly increase the number of known \cref{bright $z>3$ lensed galaxies}, as these data are deeper, particularly in redder filters, than the SDSS. Here, we report and characterize one such object, a spectroscopically confirmed $z$=5.043 strongly-lensed galaxy, with an extraordinarily bright $z_{AB}$ apparent magnitude of 20.47. As such, this is the brightest $z\gtrsim$5 UV-bright galaxy found so far.

\begin{figure*}[t]
\epsscale{1.23}
\plotone{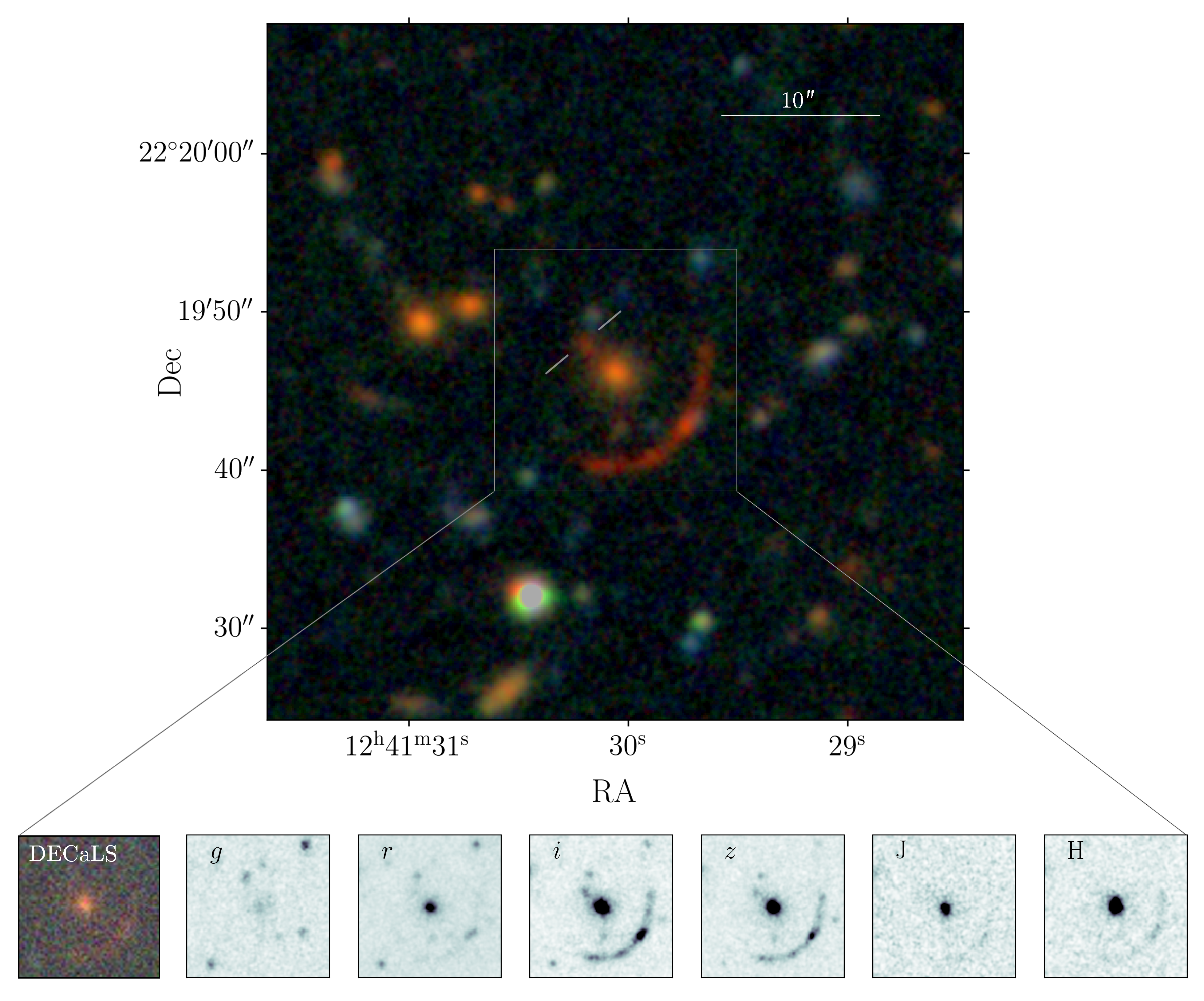}
\caption{(Top) RGB image of the lens and COOLJ1241+2219. This was constructed using the Magellan/PISCO $g,r$ and $i+z$ images in each channel. The arc radius about the central lens galaxy is $\sim$6”, indicating that the lens has the mass of a large group or small cluster, consistent with the presence of other red sequence galaxies. The faint red source immediately NW of the main lens galaxy has colors and surface brightness suggestive of a lensed counterimage, marked with solid white lines. (Inset images) $griz$JH images from a 16"x16" field-of-view centered at the lens. (Inset image: far left) The central portion over the same area of the 45"x45" DECaLS $grz$ discovery image.
\label{fig:rgb}}
\end{figure*}

This paper is structured as follows. Section \ref{sec:discovery} lays out the details of the public data search that led to the discovery of COOL-J1241+2219. Section \ref{sec:data} describes follow-up imaging and spectroscopy of COOL-J1241+2219, while Section \ref{sec:analysis} describes the analyses of these data and the results from stellar population synthesis and strong lens modeling. Section \ref{sec:discussion} places these results in the context of other objects known in the $z>5$ Universe. 

All reported magnitudes are calibrated to the AB system. The fiducial cosmology model used assumes a standard flat cold dark matter universe with a cosmological constant $(\Lambda$CDM), corresponding to WMAP9 observations (\citealp{Hinshaw_2013}). For inferred parameters with uncertainties, we report 16th, 50th and 84th percentile values, unless otherwise specified.

\section{Discovery} \label{sec:discovery}

The COOL-LAMPS project — ChicagO Optically-selected strong Lenses - Located At the Margins of Public Surveys — initiated as the central focus of an undergraduate research class, is an effort to find strong gravitational lenses in recent public imaging data. Though the lensing search is designed to find a wide variety of lenses and lensed sources, the particular focus of our follow-up is sources that are photometrically at the margins of the distributions of source color and brightness. The details and results of this search will be presented in an upcoming publication \citetext{COOL-LAMPS Collaboration in prep.}, and we provide only a brief description here.

\targ was found in the DECaLS $grz$ Data Release 8 (DR8, \citealp{Dey_2019}) imaging data, as part of a comprehensive visual search of the northern galactic cap portion of the DECaLS dataset. The complete search comprised two separate efforts, one targeting potential lenses at low redshift ($z<0.7$) and the other extending to higher redshift potential lenses. COOL-J1241+2219 was found in the latter.

\begin{figure*}[t]
\epsscale{1.20}
\plotone{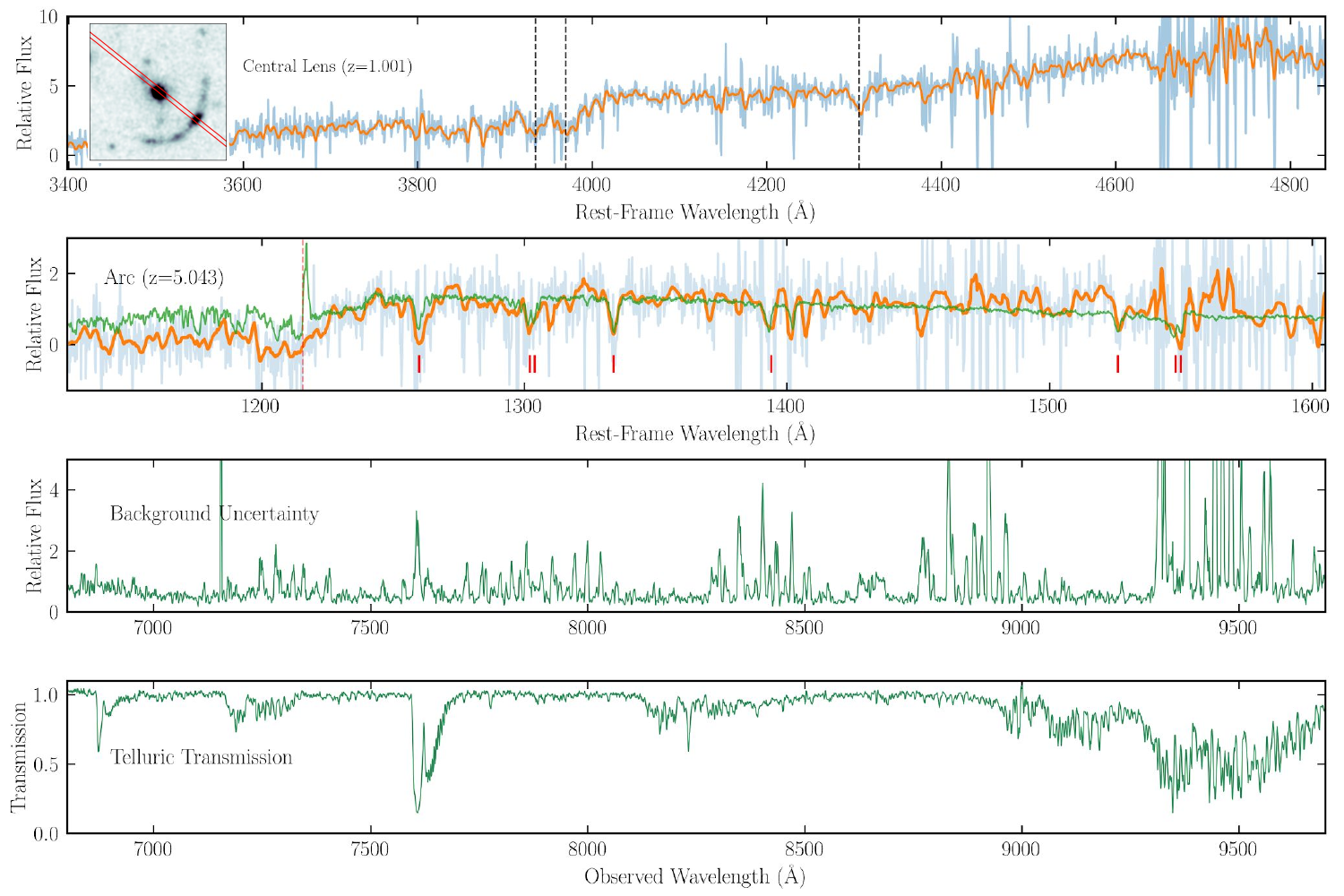}
\caption{The extracted optical/near-IR Magellan/LDSS3 spectra of the central lens galaxy (top), brightest knot in the arc (middle top), the noise spectrum (middle bottom) and telluric correction applied (bottom). The noise and transmission spectra are shown covering the same observed wavelength range, corresponding to the indicated restframe wavelength range for redshifts of 1.001$\pm$0.001 and 5.043$\pm$0.002 for the top and middle top panels respectively. The raw spectra are plotted in pale blue, and versions of those data filtered (via a median and tophat filter kernel) to visually emphasize features relevant for redshift determinations, are over plotted in orange. The inset top-left shows the slit positioning for these observations. In the top panel, vertical dotted lines indicate significant spectral features used to infer the redshifts for the lens galaxy; black lines (from left to right) mark the CaII H\&K lines, and the G-band. For the lensed source spectrum, we mark for reference Ly$\alpha$ in absorption at 1215\AA, though note that the redshift is determined from an uncertainty-weighted cross correlation with the reference spectrum (from \citealp{Rigby_2018b}) overplotted in green. To ease visual comparison, we have scaled the reference spectrum using a low-order polynomial computed redward of Ly$\alpha$ to match its amplitude to the lensed source spectrum. \cref{To visually confirm the redshift, we mark UV stellar and nebular absorption features with red lines — (from left to right, in rest-frame \AA\ units) SiII 1260, OI/SiII 1302/1304, CII 1334, SiIV 1393, SiII 1526, CIV 1548/1550.}
\label{fig:ldss3_arc}}
\end{figure*}

We selected luminous red galaxies, classified as having de Vaucouleurs profiles (\citealp{1948AnAp...11..247D}), and appropriate colors and magnitudes, as potential lenses. Color and magnitude cuts were informed by prior experience with strong lens selection in the SDSS (\citealp{Bayliss_2011,Sharon_2020}; \cref{to first-order, the selected galaxies are objects brighter and redder than a color-magnitude model corresponding to a passively evolving galaxy formed at $z$=3, with an absolute magnitude tuned to capture a reasonable number of target objects. In practice, this magnitude limit is approximately M$_{*}$. Specifically, \targ's measured photometry puts it at $\sim$ 0.2 magnitudes redder and brighter than the selection cut, and further away from the selection cut than the median potential lens selected by that cut.}

We followed a visual examination strategy employed by the Sloan Giant Arcs Survey (SGAS; e.g., \citealp{2011ApJS..193....8B,Sharon_2020}). In the high-redshift portion of our search, a total of 271,460 lines of sight were examined by at least 3 of the co-authors, primarily via a custom \code{Python}-based image viewer and lens-ranker. We examined lines of sight four-at-a-time in a 2$\times$2 grid, with each individual cutout covering $45"\times45"$ centered on each potential lens. We constructed custom color images from the DECaLS $grz$ imaging data, tuned to emphasize faint extended features. Each 2$\times$2 composite image was given an integer score from 0 to 3 by each examiner, with a score of 3 indicating a definite strong lens, a score of 0 indicating no evidence for strong lensing, and the remainder shading between those extremes. The initial average score for COOL-J1241+2219 was 2. The central portion of the image in which COOL-J1241+2219 was found is shown in the first inset panel of Figure \ref{fig:rgb}. 

The lensed arc is detected in the DECaLS catalogs at RA, Dec (J2000) = $(190.3743, 22.3282)$ with reported AB magnitudes of $g$=25.10, $r$=23.82, $z$=21.13; the model-based photometry of these catalogs assumes elliptical sources, and must therefore miss a significant fraction of the source light in this case. However, this initial photometry and the observed red color and morphology of this apparent arc motivated follow-up observations.

\vspace{10mm}

\section{Follow-up Observations and Redshifts} \label{sec:data}

\subsection{Photometry}
Near-infrared imaging of \targ in the J- and H-bands was obtained using the FourStar Infrared Camera (FOURSTAR; \citealp{Persson_2008}) on the Magellan/Baade telescope, Chile, on 2020 February 22. The total integration times were 335s and 3578s in the J- and H-bands respectively. Data were reduced to final astrometrically- and photometrically-calibrated stacked images using a custom pipeline built via IRAF and PHOTPIPE (\citealp{Rest_2005, Garg_2007, Miknaitis_2007}).

We then obtained deep simultaneous optical imaging — $griz$ — using the Parallel Imager for Southern Cosmology Observations (PISCO; \citealp{Stalder_2014}) on the Magellan/Clay telescope, on 2020 February 29. The total integration time for the imaging with a field-of-view of 5.5' x 7.5' was 3060s. Data were processed and stacked to final astrometrically-calibrated images using a custom pipeline that makes standard corrections, as well as removing effects of non-linearity and bright stars peculiar to the PISCO detectors. Figure \ref{fig:rgb} shows a color image for \targ constructed from the PISCO images, as well as smaller greyscale cutouts for all of the follow-up imaging focused on the arc and central lens galaxy.

As can be seen in Figure \ref{fig:rgb}, these data show that \targ is blue in the z—J and z—H colors. In combination with the observed red optical colors this indicates that this arc is the strongly-lensed image of a high-redshift galaxy.

\subsection{Spectroscopy}

Spectroscopic observations for \targ were obtained on 2020 March 2, using the Low Dispersion Survey Spectrograph - 3C (LDSS-3C\footnote{http://www.lco.cl/Members/gblanc/ldss-3/ldss-3-user-manual-tmp}), with the VPH-Red grism and a 1\arcsec.0 longslit placed to cover a wavelength range of 6700\AA\ and redward, to the limits of the instrument response at $\sim$10500\AA\ . The slit was oriented to include the brightest clump in the lensed arc, the fiducial counter-image, and the apparent lens galaxy (with a slight offset). The precise slit placement is shown in the top inset portion of Figure \ref{fig:ldss3_arc}. 

We acquired six 900s integrations, dithered between two slit positions along the slit diection, in non-photometric conditions at airmasses of 1.6-1.8, with a typical on-instrument seeing of 1\farcs{1}, as well as observations of a spectrophotometric standard for both flux calibration and the removal of telluric features in the spectra. Reduction to wavelength- and flux-calibrated 1D spectra was accomplished using a combination of standard routines and custom software, implemented in IRAF and IDL. The spectra for the apparent central lens galaxy and \targ are shown in Figure \ref{fig:ldss3_arc}, along with an estimate of the noise spectrum, and the applied telluric correction. We measure the lens redshift from the marked features in Figure \ref{fig:ldss3_arc} as $z=1.001\pm0.001$, and the lensed source redshift from cross-correlation with the stacked rest-UV galaxy spectrum from \cite{Rigby_2018b} as $z=5.043\pm0.002$.

The high-airmass and non-parallactic slit positioning of these observations in sub-optimal conditions makes the fluxing and telluric calibrations irretrievably uncertain. However, as is obvious from Figure \ref{fig:ldss3_arc}, both spectra are sufficient to establish redshifts robustly. 

At the measured redshift, the entire restframe optical band \cref{for the lensed galaxies} is shifted to wavelengths not readily accessed from ground-based observatories. The only observable standard emission line used in studies of galaxies at lower redshifts is the [O II] 3727,3729\AA\ doublet. We observed \targ using the Gemini-North Near-infrared Spectrograph (GNIRS) spectrograph (\citealp{Elias_2006}) at NOIRLAB/Gemini, Mauna Kea, in Director's Discretionary Time on July 4th, 2020 in an attempt to measure this feature. We acquired eight 257s integrations using GNIRS with a 0\arcsec.675 slit in cross-dispersed mode and with the short camera, at an airmass$\sim$1.5, with a typical on-instrument seeing of 0\arcsec.6. These observations targeted the same bright peak along the arc observed with LDSS-3C. The data were reduced primarily using the  Gemini/GNIRS pipeline\footnote{https://www.gemini.edu/instrumentation/gnirs/data-reduction}. The resulting spectrum was calibrated against a telluric standard star observed immediately after. Analysis of these data is detailed in Section \ref{sec:analysis}.

\vspace{5mm}

\section{Analysis and Results}\label{sec:analysis}
\subsection{Zeropoint Calibration}

 The J- and H-band images are calibrated to 2MASS stars (\citealp{Jarrett_2000}) within the field of view, with the calibration derived automatically by PHOTPIPE routines. Uncertainties on these zeropoints relative to 2MASS are $\sim$ 0.03 magnitudes.
 
 The PISCO data are calibrated to existing DECaLS ($grz$-bands) and Pan-STARRS (DR2,\citealp{chambers2016panstarrs1}) ($i$-band) data. We used SExtractor (\citealp{Bertin_1996}) MAG\_AUTO magnitudes for detected objects in all bands to establish image zeropoints. The PISCO data are much deeper than the reference data, and so object-by-object uncertainties in the comparison are dominated by uncertainties in the reference measurements. Limiting the comparison to objects with a measured DECaLS SNR of $>$20 yields a matched catalog sufficient to established zeropoints, and linear color terms, accurate to better than 0.02 magnitudes, and we derive a similar result in comparison to the Pan-STARRS data.

\begin{figure*}[t!]
\includegraphics[scale=0.6]{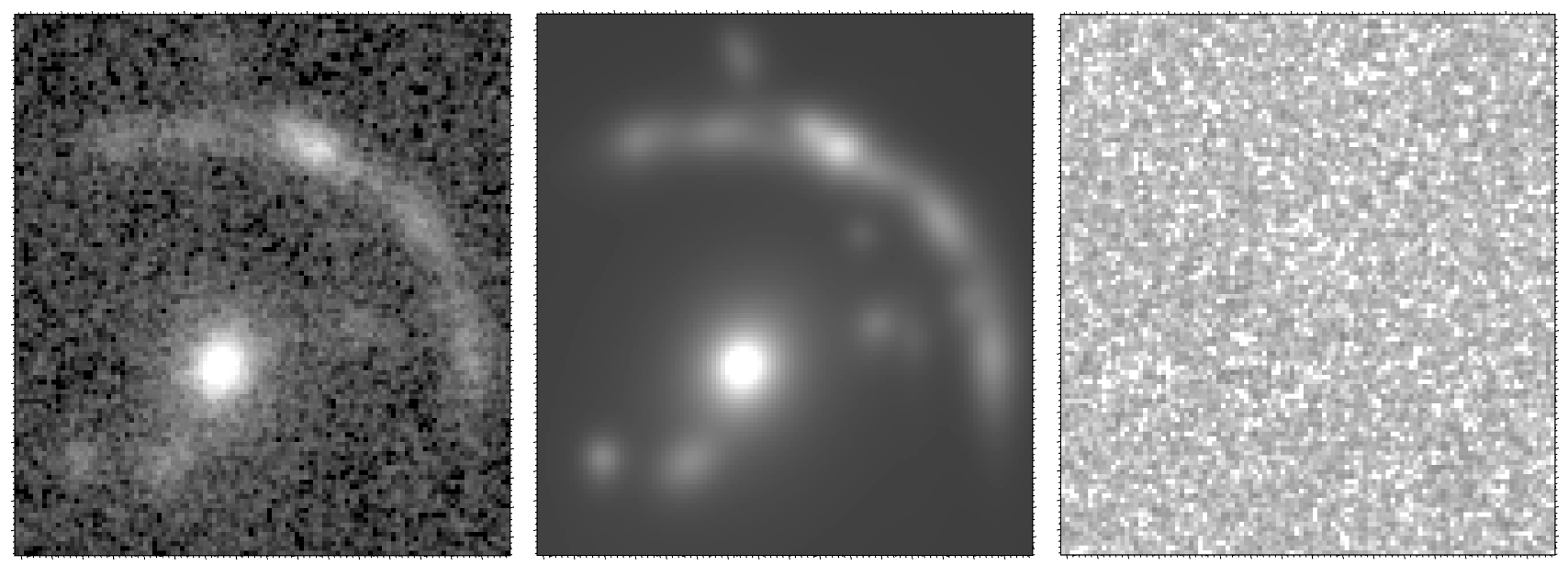} 
\label{fig:i_pisco_galfit}
\caption{12"x12" cutouts, all at the same intensity scaling, of the (left) PISCO \textit{i} band image, (center) GALFIT model of the central lens galaxy, the arc, and other galaxies, and (right) the residual image (with inverted colors to highlight the lack of residual structures).}
\end{figure*}
\newcommand{\datacaption}{Model photometry-based magnitudes for \targ}
\newcommand{\datacomments}{Magnitudes calculated via GALFIT, with output in AB system. $griz$-band imaging from Magellan/PISCO simultaneous multi-band imager. JH band imaging from Magellan/FOURSTAR infrared imager.}
\begin{deluxetable*}{l ccccccc}
\tablecolumns{7}
\tablewidth{0pt}
\tablecaption{\datacaption}
\tablehead{& $g$ & $r$ & $i$ & $z$ & J & H}
\startdata
\cref{Central Lens Galaxy} & 23.86 $_{-0.20}^{+0.24}$ & 
21.960$ \pm 0.046$ & 
20.76$ \pm 0.05$ & 
19.630$ \pm 0.048$ & 
18.955$ \pm 0.059$ & 
18.564$ \pm 0.054$\\
Arc & >25.11 & 23.197 $_{-0.358}^{+0.353}$ & 21.616 $_{-0.050}^{+0.049}$ & 20.472 $_{-0.048}^{+0.047}$ &  20.53 $_{-0.285}^{+0.273}$ &  20.556 $_{-0.087}^{+0.084}$\\
Counterimage & — & — & 24.164 $_{-0.297}^{+0.233}$ & 22.705 $_{-0.127}^{+0.105}$ &  23.074 $_{-0.793}^{+0.453}$ &  23.774 $_{-1.545}^{+0.613}$\\
\enddata
{\footnotesize \tablecomments{ \datacomments }}
\label{table:data}
\end{deluxetable*}




\subsection{Model Photometry with GALFIT}

We used the parametric fitting code GALFIT (\citealp{Peng_2002,Peng_2010}) to create a model of the arc, its candidate counterimage, the central lens galaxy, and other nearby contaminating objects, as well as to extract relevant photometry. Point-spread functions (PSFs) were measured directly from the images; these are simply small cutouts of two isolated and bright but unsaturated point sources near the modeled region. Differences in photometry from choices of PSF are substantially smaller than all uncertainties below. Multiple two-dimensional S\'ersic components were used to model each image. Model construction proceeded iteratively, with bright objects fit first, and further components added until the residual was consistent with the background noise. 

To estimate statistical photometric uncertainties we added the final fitted model to blank sky regions around the image, and re-fit the models on these new images. The measured magnitude distribution of each object then gives an uncertainty. If the flux distribution had significant outliers (presumably due to poor convergence of the model instance) we computed the uncertainty from the flux distribution iteratively clipped at 3$\sigma$.
Additionally, to estimate the systematic uncertainty caused by modeling choices, three co-authors independently created models for the $z$-band data, and the range in final fitted magnitudes for these models was taken as an estimate of the systematic uncertainty for the PISCO images. Additionally these models were re-optimized to fit the H-band image, and provide an estimate of the systematic uncertainty in the J- and H-bands.

Finally, calibration zeropoint uncertainties, and the statistical and systematic uncertainties estimated as above, were combined in quadrature to compute total uncertainties for each measurement of each physical object.

The \textit{i}-filter image, GALFIT model and residuals for the lens field are shown in Figure \ref{fig:i_pisco_galfit}. Table 1 gives the model magnitudes for the arc and the candidate counter-image.

\subsubsection{Lens Galaxies} \label{sec:lens_gals1}
As part of the modeling, we have measured photometry for the central lens galaxy, as well as two other nearby galaxies to the east that have similar apparent colors and morphologies (and are visible in Figure 1). These two other galaxies were fit together, but independently of other objects. These two galaxies have colors similar to the apparent central lens galaxy, as is indicated also by the DECaLS photometry of these galaxies. Independent of strong-lensing mass modeling, this suggests that the lens is a lower-mass galaxy cluster. 

\subsubsection{Arc and Possible Counterimage}
The lensed source is absent in the $g$-band and nearly so in the $r$-band, both due to absorption from the intergalactic medium (IGM). It is not possible to fit a model directly in the $g$-band. Rather, we took the initial $i$-band model, solved for any residual astrometric shift between the two bands using visible objects in the modeled region, and then performed a constrained refit of the model allowing only the magnitudes to vary of any components that describe the arc. This model was pushed through the same statistical error analysis as other filters, though in the case of the $g$-band image the distribution of notional measured magnitudes represents an upper limit, and we report the 95th percentile value.

In the $r$-, J-, and H-band data, the arc is visible but measured at a markedly lower SNR than the $i$- and $z$-bands. In each of these cases we explored both direct modeling, and re-optimized models taken from bands with larger SNR detections of the arc, and the measurements we report come from the latter. Note that in all of these cases, including the $g$-band, components which describe well-detected objects were allowed to vary structurally, as well as in brightness.

The suggested counter-image (see Figure \ref{fig:rgb}) is photometrically consistent with the arc; it has consistent colors across filters where both objects are visible, and the apparent surface brightness of this object matches much of the arc. The candidate counter image was observed via LDSS-3C optical long slit spectroscopy (and not with GNIRS), but given that the lensed source has no emission lines in the rest-frame UV, we have not attempted to extract a spectrum of this much fainter source. We note, however, that there is no evidence in the 2D spectral image of any emission lines from this object that would be inconsistent with the identification of it as a lensed counterimage.

\begin{figure*}[t]
\epsscale{1.23}
\plotone{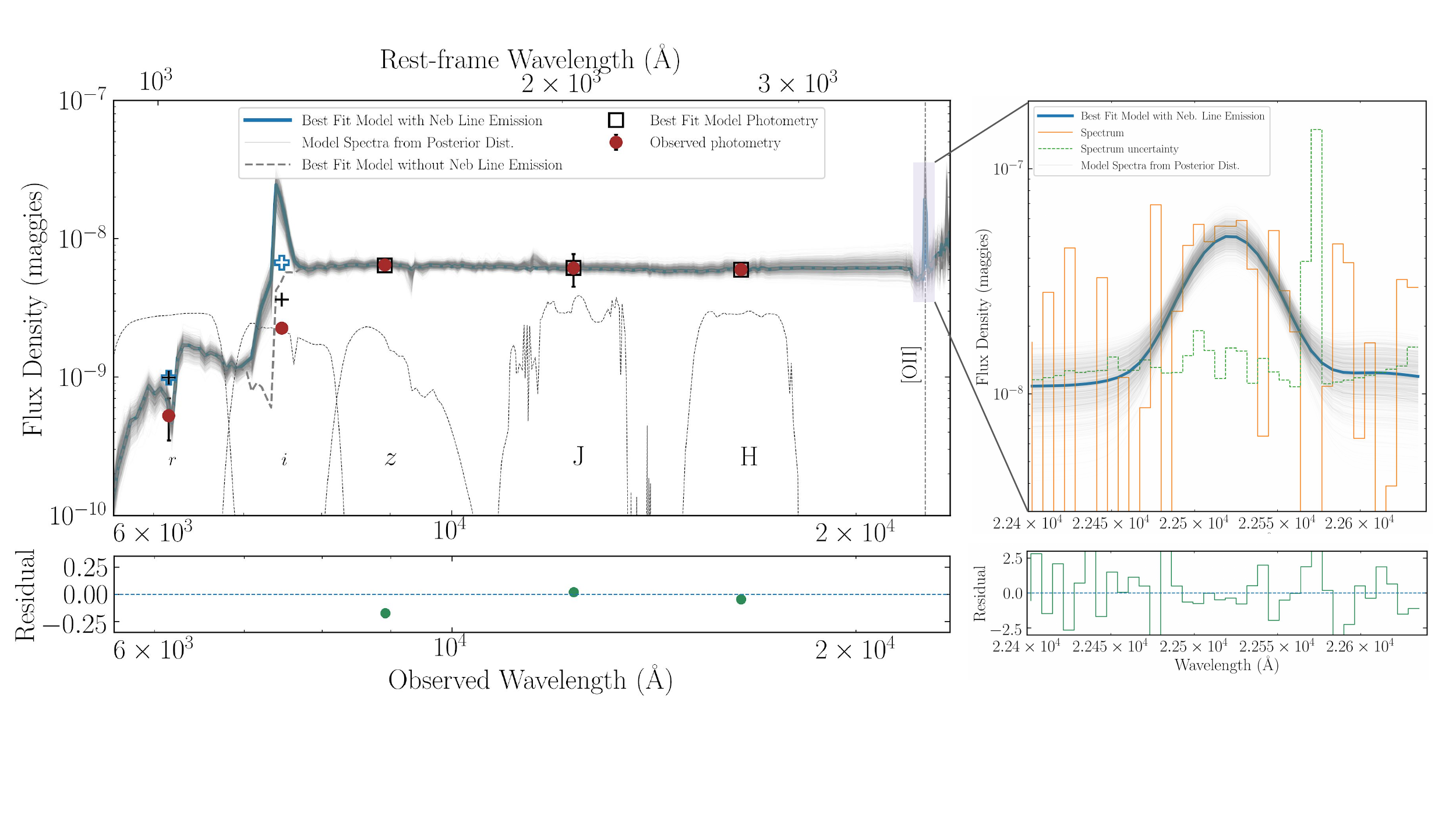}
\caption{(Left) Best fit SED model with nebular line emission (blue) and corresponding Residual ($\chi$) values (green) for \targ, calculated via \code{Prospector}, and using $z$JH photometry and IR spectroscopy. Best-fit SED model without nebular line emission is displayed as short-dashed grey curve. Best-fit photometry for $z$JH data is shown as black squares. Photometric projections for the masked $ri$ data corresponding to the model with and without nebular line emission are shown as blue 'plus' and  black 'plus' symbols, respectively. (Right) Best-fit model spectrum with nebular line emission (blue) zoomed into the observed wavelength range corresponding to \cref{the blended [O II] 3727, 3729\AA\ emission, with IR spectroscopic data (orange) and uncertainties (dotted green), with model $\chi$ values (in green) displayed in the bottom panel. Two black vertical lines mark the locations of the [O II] 3727,3729\AA\ doublet peaks.}
\label{fig:arc_bestfit}}
\end{figure*}

\begin{figure*}[t]
\epsscale{1.2}
\plotone{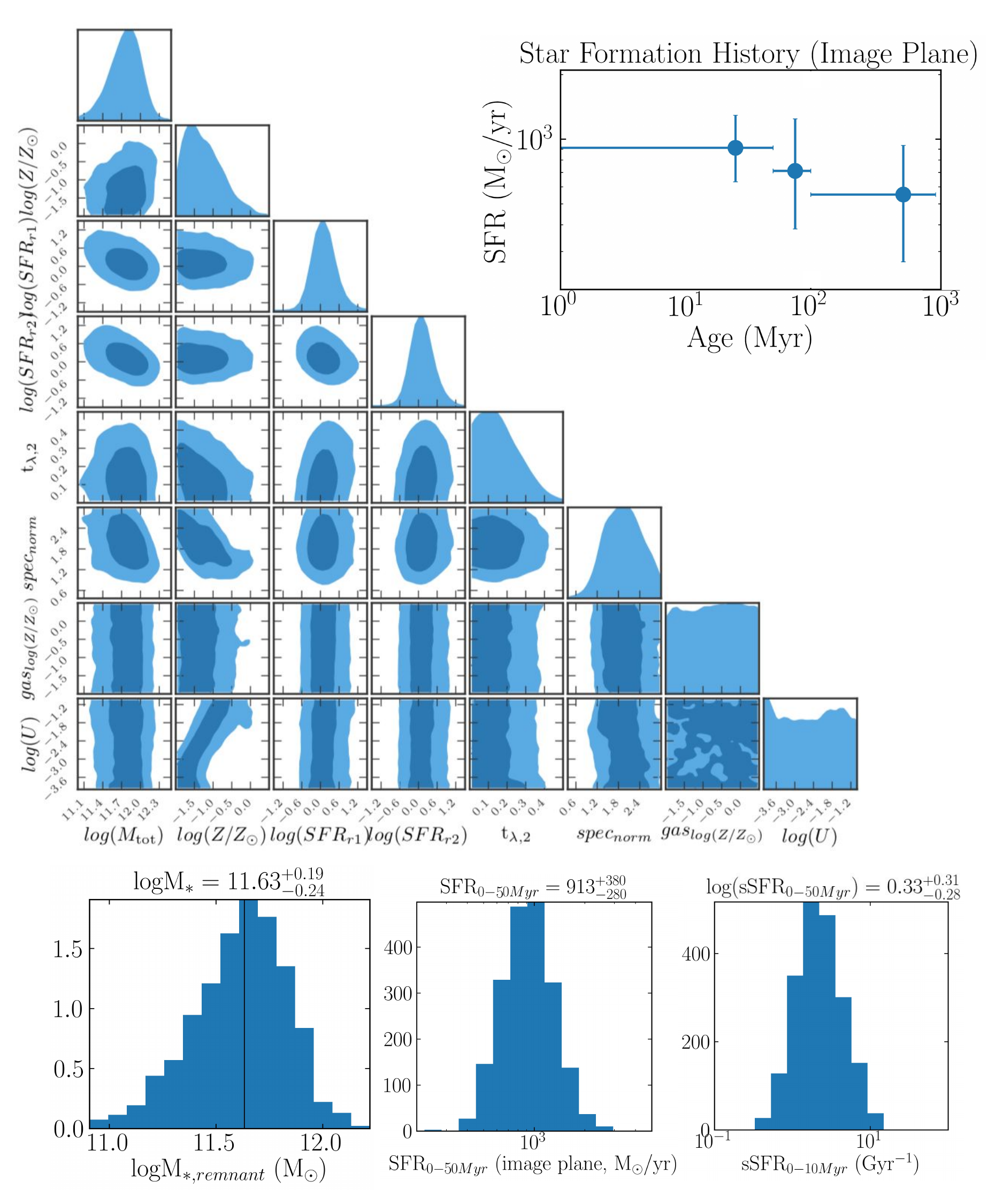}
\caption{(Top left) Corner plot with posterior distributions and correlations for \cref{inferred parameters} in the \code{Prospector} SED fitting analysis \cref{for the lensed source}, with contours corresponding to 1$\sigma$ (in dark blue) and 2$\sigma$ (in light blue). (Top right) The star formation history i.e. the SFR as a function of age bins; the uncertainties in the SFR as a function of age are calculated by Monte Carlo (MC)-sampling the SFR in the SED fitting chains. (Bottom) The posterior probability distributions of the remnant stellar mass of \targ, the star formation rate (SFR) in the youngest age bin, and \cref{the specific star formation rate (sSFR) in the same age bin}.
\label{fig:arc_corner}}
\end{figure*}

\newcommand{\arccaption}{\code{Prospector} Analysis: Free Parameters in Galaxy SED Model}
\newcommand{\arccomments}{Data in AB magnitudes. $griz$ band imaging from Magellan/PISCO simultaneous multi-band imager. JH band imaging from Magellan/FOURSTAR infrared imager.  }
\begin{deluxetable*}{l cccc}
\tablecolumns{4}
\tablewidth{0pt}
\tablecaption{\arccaption}
\tablehead{Parameter & Description & Priors & Best Fit Value}
\startdata
$\log(M_{\rm tot}/M_{\odot})$ & \cref{Total stellar mass formed} & Tophat: [10, 13] & $11.77_{-0.25}^{+0.20}$  \\
$\log(\mathrm{SFR}_{\rm ratio})$ & Ratio of the SFRs in adjacent age bins: & Student-T: Bin 1: mean=0.0,scale=0.3, $\nu=2$ & $0.10_{-0.26}^{+0.33}$ \\
& non-parametric SFH & Bin 2: mean=0.0,scale=0.3,$\nu=2$ & $0.10_{-0.28}^{+0.34}$ \\
$\log(Z/Z_{\odot})$ & Stellar metallicity in units of $\log(Z/Z_{\odot})$ & Tophat: [-2, 0.5] & $-1.32_{-0.40}^{+0.56}$ \\
$\tau_{\lambda, 2}$ & Diffuse dust optical depth & Tophat: [0.01, 1.00] & $0.14_{-0.09}^{+0.13}$ \\
$\mathrm{spec}_{\rm norm}^{*}$ & Factor by which to scale the spectrum to match photometry & Tophat: [0.1, 3.0] & $1.94_{-0.51}^{+0.53}$ \\
$\mathrm{gas}_{\log(Z/Z_{\odot})}^{*}$ & Gas-phase metallicity in units of $\log(Z/Z_{\odot})$ & Tophat: [-2, 0.5]; depends on $\log(Z/Z_{\odot})$ & $-0.94_{-0.73}^{+0.90}$\\
$\log(U)^{*}$ & Gas Ionization Parameter $U=n\gamma/n_H$ & Tophat: [-4.0, -1.0] & $-2.48_{-1.09}^{+1.00}$\\
\enddata
{\footnotesize \tablecomments{ \arccomments *Considered as nuisance parameters.}}
\label{table:sed_arc}
\end{deluxetable*}




\subsection{SED Fitting with \code{Prospector}}

Using the model photometry in $griz$JH filters, we perform spectral energy distribution (SED) fitting for the central lens galaxy and the \targ arc using the Markov Chain Monte Carlo (MCMC) — based stellar population synthesis (SPS) and parameter inference code, \code{Prospector}. \code{Prospector} is based on the the Python-FSPS framework, with the MILES stellar spectral library and the MIST set of isochrones (\citealp{Conroy_2010,prospector,Leja_2017,2013PASP..125..306F,2011A&A...532A..95F,Choi_2016}). We exclude the LDSS3 spectrum from the fitting procedure because at its low signal-to-noise ratio, it mostly samples saturated interstellar absorption lines that are not directly constrained by \code{Prospector} libraries.

\cref{We have chosen to perform the SED fitting on the observed photometry, a choice which assumes that there are no correlations within the SED fitting that couple parameters that are magnification-dependent (such as stellar mass, star formation rate) with those that are not magnification-dependent (such as metallicity, dust extinction). This assumption has been tested by arbitrarily rescaling the input photometry by the measured mean magnification (see Section} \ref{sec:lensmag}),\cref{and the resulting SED fits are entirely consistent with the results presented below.}

In these models, we assume a non-parametric star formation history (with age bins with $[0-50]$, $[50-100]$, and $[100-1000]$ Myr in lookback time), represented by the parameters SFR$_{ratio}$, referring to the ratio of total star formation in adjacent time bins. These bins correspond to redshift bins [5.04-5.22], [5.22-5.41] and [5.41- 20.50]. The priors for the star-formation rate ratios in adjacent time-bins is the continuity prior, that fits directly for the change in log(SFR) between the bins and weights against sharp transitions in SFR (see \citealp{Leja_2019}). 

We also fit for dust attenuation — using the \cite{Calzetti_2000} attenuation — applied to all light from the galaxy (in units of opacity at 5500\AA), stellar metallicity $log(Z/Z_{\odot})$ (where $Z_{\odot}=0.0142$), gas ionization parameter $U$, gas-phase metallicity (linked to stellar metallicity) and total mass formed in the galaxy (M$_{tot}$,in units of M$_{\odot}$), as free parameters. \cref{For each chain in the MCMC analysis, the remnant stellar mass} M$_{*,remnant}$ \cref{after compensating for mass loss from post main sequence stars, and stellar mass locked in stellar remnants}. The dust extinction and metallicities have flat and liberal priors, roughly covering the range allowed by the spectral model libraries. These models each assumed a Kroupa IMF \citep{Kroupa_2001}. Nebular continuum and line emission are present, and a nominal velocity smoothing of 250 kms$^{-1}$ was used for the spectrum, informed by and consistent with both an SED fit with this as a free parameter as well as  a direct fit of the spectrum using a simple Gaussian (see below).

The arc is not visible in $g$, and the $r$- and $i$-bands sample the Ly$\alpha$ forest and Ly$\alpha$ and Ly$\beta$ in the arc. To avoid biases arising from a lack of correlation between star formation activity and the complexity of Ly$\alpha$ absorption/emission in galaxies, which is not captured in the stellar population synthesis models, we use only $z$JH photometry for the arc SED fitting. We add to this the IR spectroscopy sampling the [O II] 3727,3729\AA\ doublet. The spectrum and photometry are fit here simultaneously. 

We assume that the [O II] equivalent width is uniform across \targ in the source plane, and we scale the emission line spectrum to be representative of the whole arc by considering the entire arc flux against the portion contained within the slit, in the $i-$band. This flux normalization is uncertain in practice, quite apart from the underlying assumption of uniform equivalent width,  so we include a nuisance parameter, spec$_{norm}$ (spectrum normalization factor), to capture any unresolved differences between model photometry and spectroscopy. This parameter is simultaneously fit with other parameters within \code{Prospector}. Note however, that fits computed without this parameter in place are consistent with conclusions from models (e.g., stellar mass) that include spec$_{norm}$. 

The best fit SED model is shown in Figure \ref{fig:arc_bestfit}. We also show corner plots constructed via \code{pyGTC} (\citealp{Bocquet2016}), with one and two dimensional projections of the posterior probability functions (to demonstrate both one dimensional marginalized probabilities and two dimensional covariance) in Figure \ref{fig:arc_corner} for the free parameters fit in the arc SED modeling (see Table \ref{table:sed_arc}).

It is also crucial to note that the best fit SED model fits a nominal Gaussian to the observed wavelength range where we expect [O II] emission. The result here -- mean, width and amplitude -- of the [O II] emission feature, is consistent with results from fitting a 1D Gaussian profile to just the spectroscopic data (see Section \ref{sec:o2} for details).

The parameter M$_{tot}$, total stellar mass formed, converges to a tailed Gaussian posterior distribution for these sources, which corresponds to a model-generated remnant stellar mass distribution, displayed in the bottom panel of Figure \ref{fig:arc_corner}. The median value for the remnant stellar mass (M$_{*}$) in the image plane (i.e., the magnified value) is logM (M$_{\odot}$)) = $11.63^{+0.19}_{-0.24}$. The precision of this result may surprise the reader, given that it is based on rest-UV photometry, which is under most circumstances — e.g. at lower redshifts, and in the presence of significant extinction — insufficient to constrain stellar mass reliably. However, only modest dust columns are allowed by the data, and at $z=5.043$ the Universe was not old enough to conceal much stellar mass in older stars that do not produce significant UV flux.

We further test different models to check the robustness of our results. The stellar mass of the arc is found to be robust when varying parameters as follows:\\
1.Stellar metallicity fixed at the median best fit value in Figure \ref{fig:arc_corner}.\\
2. Removing nebular line emission from the model.\\
3. Altering the number of age bins from three to four, which is recommended by \citep{Leja_2019} (as well as from three to two), spanning the same time period. \\
4. Changing mean and standard deviation of the star formation history priors.\\
5. Assuming a parametric star formation history (decayed-tau with a single burst).\\
6. Constraining SFH only to 0-500 Myr, as opposed to 0-1000 Myr. \\
7. Using velocity smoothing as a free parameter for the [O II] emission line fitting. \\
8. Fixing gas ionization parameter to -2 and -2.5, as well as gas-phase metallicity to that of the stellar value (i.e. 0). \\

We use the Bayesian Information Criterion (BIC) to quantify our model selection, for the greater penalty imposed by BIC on the number of free parameters used in our fitting procedure. Given a candidate model and M models to compare with, the magnitude of $\Delta$BIC (BIC$_{M}$ — BIC$_{candidate}$) can be inferred as evidence against a candidate model being the best model \citep{kass1995bayes}. The candidate best-fit model has a BIC value of -1245.5, and the models discussed above have a |$\Delta$BIC| $<$ 5, which implies that the evidence against our candidate best-fit model is positive, albeit low. Qualitatively, all models constrain stellar mass and dust attenuation, while leaving metallicities and the gas ionization parameter unconstrained (albeit used here as nuisance parameters).

It is important to note that the SED fitting model contains contribution by nebular line emission in order to fit [O II] emission, and hence depicts Ly$\alpha$ in emission even though it is absent in the spectrum and photometry (as indicated by mismatch between observed photometry and the synthetic magnitudes marked by plus symbol in Figure \ref{fig:arc_bestfit}). This does not impact our results since the photometry sampling Ly$\alpha$ is not included in our candidate best-fit model, and M$_{*}$ derived from removing nebular emission contribution is within uncertainties consistent with that derived from our best-fit model.

To cross-check the complete end-to-end pipeline from raw data to SED fits, we have also applied a \code{Prospector}-based analysis to measurements of the central lens galaxy and two likely cluster members discussed in Section \ref{sec:lens_gals1} above, and find that these galaxies are well fit by passively aging old stellar populations at the notional lens redshift — as expected.

\subsection{Instantaneous SFR: [O II]}\label{sec:o2}

Distinct from the simultaneous spectro-photometric SED fit to the data, we cross-check our SFR result by calculating the [O II] 3727,3729\AA\ doublet line flux to characterize the instantaneous SFR \citep{2002MNRAS.332..283R} for \targ from a scaling relation. We fit a 4 parameter 1-D Gaussian profile (with mean, width, amplitude and continuum pedestal) to the spectroscopic data, in the rest-frame region 3720-3735\AA, to capture the blended-doublet emission line. The observed [O II] line flux, corrected for the entire arc, is $4.5^{+2.1}_{-2.0} \times 10^{-16}\ ergs cm^{-2} s^{-1}$. \cref{The redshift derived from blended-doublet line is $5.041 \pm 0.001$}. 

Using the oft-used Kennicutt-calibrated [O II]-SFR relation (\citealp{Kennicutt_1992}):
\begin{equation}
SFR ([O II], M_{\odot}yr^{-1}) = 1.4 \times 10^{-41}  L_{[O II]}(erg s^{-1})
\end{equation}

\noindent in the image plane we find the SFR for \targ to be $1700^{+800}_{-800}$ M$_{\odot}$ yr$^{-1}$. Instead, if we use the SFR-[O II] relation empirically corrected for stellar metallicity and takes as input the galaxy stellar mass (Equation 8 in \citealp{Gilbank_2010}), we find the calculated SFR to be $4600^{+3400}_{-2400}$ M$_{\odot}$ yr$^{-1}$.

Without stronger constraints on metallicity and dust in our SED posterior distributions, their impact on the systematic uncertainties in the SFR-[O II] flux scaling relations cannot be mitigated (e.g., \citealp{Kewley_2004,Bicker_2005}). It is important to note that current constraints are within the metallicity values observed for local universe galaxies.

\begin{figure}[t]
\epsscale{1.2}
\plotone{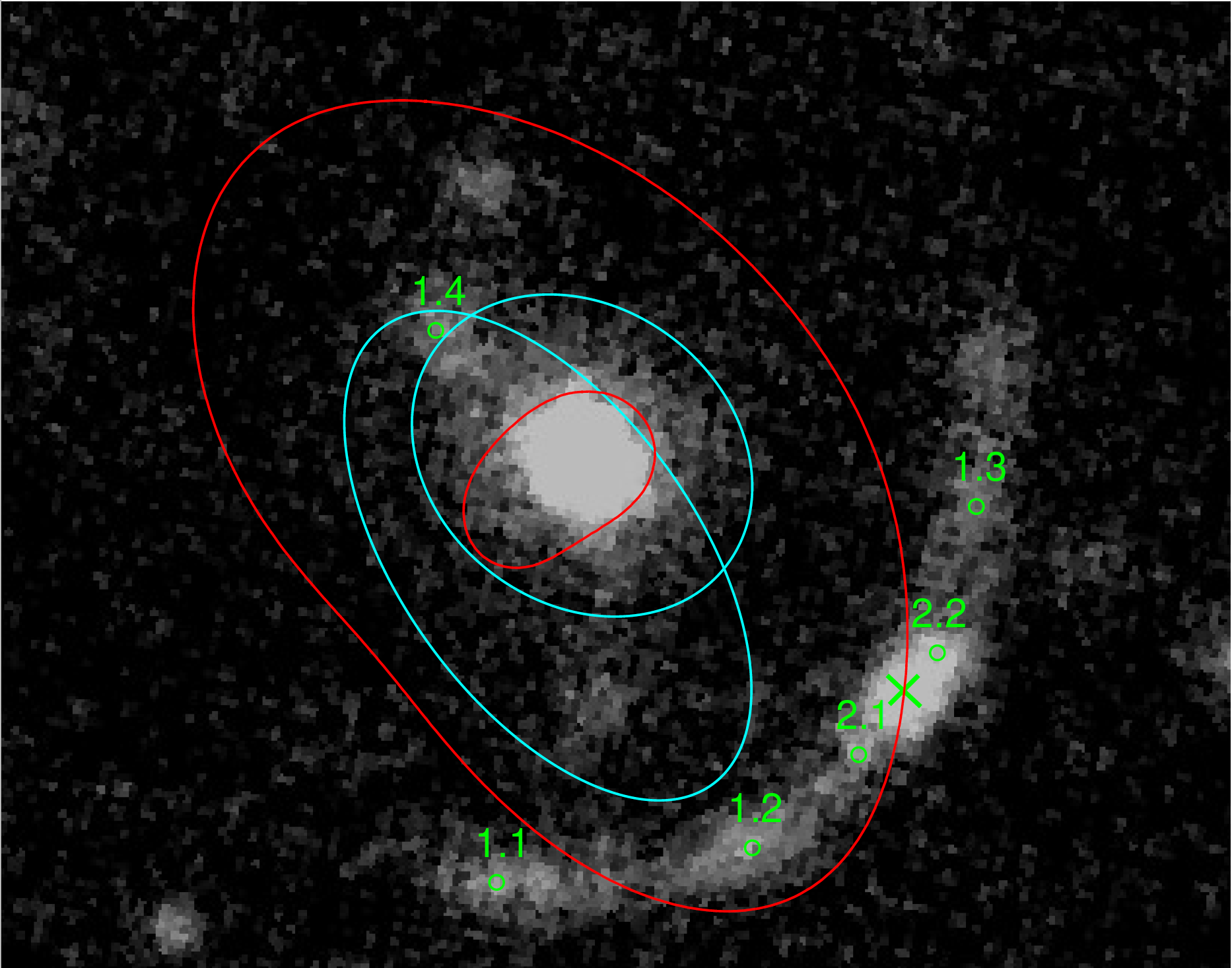}
\caption{Lens model of the low-mass cluster lens based on ground-based imaging, using the positional constraints indicated in green. We include two halos — a galaxy-scale (smaller cyan ellipse) and cluster-scale halo (larger cyan ellipse) to explain the observed lensing. The size of the ellipses represent, on an arbitrary scale, the relative masses of the halos. The tangential critical curve in red trisects the arc – this is a classic 3-image arc with a counter image. \cref{This mass configuration is also informed by the bright knot along the arc, which we set as a critical curve crossing in the model (green cross)}. 
\label{fig:lens_model}}
\end{figure}

\begin{figure*}[t]
\epsscale{1.0}
\plotone{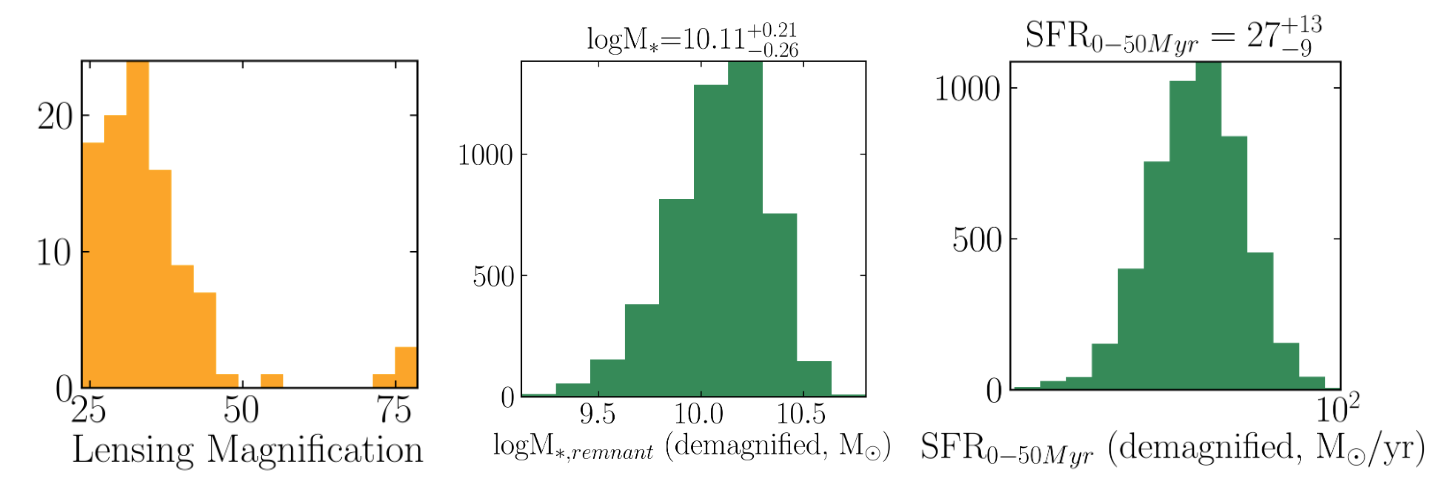}
\caption{(Left) Distribution of magnification values, from a sample of 1000 magnification maps generated via an MCMC LENSTOOL analysis of the lens model fit to the data. Note that distribution is strongly asymmetric about the best fit value, which is in part a consequence of the choice to identify the brightest peak on the arc as a location of the critical curve. (Right) Current SFR in the source plane.
\label{fig:mag_dist}}
\end{figure*}

\subsection{Lens Modeling and Magnification}\label{sec:lensmag}

We provide a brief summary of the gravitational lensing analysis used in this work here, and we refer the reader to \citealp{Kneib_1996,Richard_2010,Verdugo_2011} for a more in depth discussion of the lensing algorithm used. We adopt a parametric approach using \code{LENSTOOL} (\citealp{Jullo_2007}) to model the mass distribution with dual pseudo-isothermal ellipsoids (dPIEs, \citealp{2007arXiv0710.5636E}), using an MCMC method to estimate the parameters and their uncertainties. The lens modeling strategy generally follows \cite{Sharon_2020} (with a cosmology with $\Omega_M$ = 0.3, $\Omega_{\Lambda}$ = 0.7). We model the lens plane with two dPIE potentials, one to represent a group-scale dark matter halo, and one that represents the contribution from the main lens galaxy. The geometric parameters of the galaxy-scale halo are fixed to its observed properties (position, ellipticity, and position angle), while the slope and overall normalization are allowed to vary. The choice of two halos is motivated by the arc radius, which suggests the presence of a mass structure larger than a single massive galaxy, and the presence of other luminous early-type galaxies with colors consistent with the apparent central lens galaxy, again indicative of a more complex structure. Hence, the two dPIE clumps used here are combined to map the dark matter (DM) at the cluster scale and the central lens galaxy. 

The smooth appearance of the arc in the ground based data limits the availability of lensing constraints; a more complex lens plane, with additional galaxy-scale halos, is not required in order to satisfy these constraints. Hence, we do not include any other galaxy-scale halos in this initial model (see Figure \ref{fig:lens_model}). \cref{We choose to use the peak of the galaxy core as a constraint (see system 1 with four images in Figure \ref{fig:lens_model}), the surroundings of the star-forming clump merging on the critical curve (system 2), and a critical curve constraint (see green cross). The precision assumed for the location of the constraints in the lens model is 0.1", with an RMS of the optimized model in the image plane of 0.12". }

To estimate the magnification, we calculate the ratio of flux in the image plane using the GALFIT modeling image of the arc and its pixel-matched \cref{source plane reconstruction} for 100 realisations of the lens model posterior probability distribution, resulting in an estimate of the magnification of $32^{+8}_{-5}$. We apply this as a division factor to the arc imaging in the image plane, to convert the derived physical parameter values to the source plane. 

The uncertainties used in the SED fitting do not include the system-wide magnification uncertainty, which when combined with the statistical uncertainties, yields a final stellar mass for \targ of logM (M$_{\odot}$) = $10.11^{+0.21}_{-0.26}$. Similarly, the demagnified current star formation rate (SFR, in units of M$_{\odot}$/yr) is calculated to be $27^{+13}_{-9}$ (see Figure \ref{fig:mag_dist}), which makes this an active star-forming galaxy, given the constraints of the SED and lens modeling. This is assuming that the SFR in the age bin 0-50 Myr is a proxy for current star forming activity in \targ.

\vspace{5mm}

\section{Discussion}\label{sec:discussion}

\subsection{Luminosity Function}

We compare the brightness of \targ to M$^*_{UV}$, the characteristic luminosity in the parametric luminosity function \citep{1976ApJ...203..297S}. As a proxy for the UV absolute magnitude we compute $M_{1600}$, calculated from synthetic photometry applied to the  best fit SED model, using a top-hat filter from 1550-1650 \AA. The UV continuum is roughly flat, and hence differences in definitions of UV magnitudes in the literature do not significantly impact the comparison of our reported absolute magnitude to published values.

Taking into account the lensing magnification, M$_{UV}$ for \targ is $-22.2^{+0.2}_{-0.2}$. \cite{Bouwens_2015} give a characteristic M$^*_{UV}$ at $z$ = 5 of -21.1$\pm$0.2 (see Table 6 in \citealp{Bouwens_2015}; we use the higher uncertainty value). Hence, we find that for \targ, M$^*_{UV}$ — M$_{UV}$= $1.0^{+0.5}_{-0.4}$. In other words, \targ is $\sim$ 2-4 times more luminous than an object with the characteristic UV luminosity at these redshifts. 

\subsection{The UV Continuum Slope}
The observed UV continuum slope of star-forming galaxies is closely related to emission from massive stellar populations, as well as dust extinction in the galaxy. We fit the best-fit SED model for \targ with a power law in f$_{\lambda}$, i.e., f$_{\lambda}$ $\simeq$ ${\lambda}^{\beta}$, between rest-frame 1268-2580\AA\ \citep[the entirety of the fitting windows described initially by][]{Calzetti_1994}. For \targ, the best-fit UV continuum slope is calculated to be $\beta$ = -2.2$\pm$0.2. This is on the blue side of the distribution of $\beta$ observed in (albeit fainter) galaxies at $z=5$ in \citep[e.g.,][]{Finkelstein_2012} but consistent with their measured distribution.  However, the mean relation between $\beta$ and UV luminosity \citep{Bouwens_2014} for galaxies at $z=5$ predicts a value of $\beta$ for \targ of $\sim$-1.6, significantly shallower than we observe. The observed $\beta$ is in fact consistent with that for galaxies $\sim$ 4 magnitudes fainter, emphasizing that \targ is unusually blue for its total UV luminosity.

\subsection{Star-forming Main Sequence}

The relationship between SFR and stellar mass of a galaxy is a well-studied relationship in galaxy populations and large samples, known as the star-forming main sequence (SFMS; \citealp{Noeske_2007,Rodighiero_2011}). It is informative to place these properties of \targ in relation to the SFMS of a galaxy sample in the same epoch of observation. We compare the SFR and stellar mass of \targ with galaxies at 4 < $z$ < 6 in the Hubble Space Telescope Frontier Fields. The SFMS for galaxies in this field in the redshift bins 4 < $z$ < 5, and 5 < $z$ < 6 has been reported in \cite{Santini_2017}. We direct the reader to the solid lines in bottom panels in Figure 2 of that publication, which reflect the best-fit SFMS, corrected for Eddington bias. The demagnified SFR and stellar mass of \targ (logM (M$_{\odot}$) = $10.11^{+0.21}_{-0.26}$ and SFR (M$_{\odot}$/yr) = $27^{+13}_{-9}$), are consistent with these power-law fits for the SFMS within 1$\sigma$, but incompatible with high SFRs corresponding to bursty mode(s) of star formation during major mergers or in the densest regions. It is also important to note that the source plane properties of \targ place this galaxy among the highest mass values reported in those data.

\subsection{Stellar Populations in \targ}

\targ has a stellar mass consistent within a factor of a few of that of the present-day Milky Way \citep{2015ApJ...806...96L}, and is forming stars $\sim$ 1-2 orders of magnitude faster. This rate of star formation is not surprising, given that the available time to build this galaxy at $z=5.04$ is only $\sim$1 Gyr. \targ is extremely UV-bright with little apparent extinction. This stands in strong contrast to some similarly massive galaxies known from this early epoch \citep[e.g.,][]{Riechers_2013,Strandet_2017,Casey_2019} which tend to be much dustier with SFRs an order of magnitude larger. This difference is of course due to the selection process used to identify these objects. What is less clear, however, is whether \targ looked like these dust-enshrouded examples of early star formation at any point in its history. The relative contribution of dust-enshrouded star formation to the total star formation rate at this epoch is not well-measured \citep{Casey_2019} and so objects like \targ that are bright enough to \cref{allow detailed studies} are of particular interest.

The current constraints from SED fitting (see Figures \ref{fig:arc_corner} and \ref{fig:mag_dist}) imply that the SFH of \targ is consistent with a constant SFR across the $\sim$1 Gyr available. The current data however are insufficient to say more; the non-parametric model forms the bulk of its stars in the oldest (and widest) bin, and a briefer and much elevated star-formation episode in that interval is certainly possible.

Despite significant ongoing star-formation, and a blue UV slope that suggests little extinction, \targ shows no evidence for Ly$\alpha$ emission \cref{in the extracted spectrum. In addition, we carefully examine the stacked 2D spectrum, and see no evidence of spatially-extended Ly$\alpha$ emission outside of the continuum. This complete lack of emission} indicates the presence of sufficient dust to suppress Ly$\alpha$, as one would expect in a massive galaxy with a sustained SFH and resulting enrichment. These results are somewhat in tension. However, note that the measured scatter in galaxies between the Ly$\alpha$ escape fraction and continuum extinction is large \citep{Hayes_2011} and suppression of Ly$\alpha$ emission is possible even with extremely sparse dust columns.

\section{Summary and Future Work}

We report the discovery of \targ, a strongly lensed galaxy at $z$=5.043$\pm$0.002 that is lensed by a moderate-mass galaxy cluster at $z$=1.00$\pm$0.001. In this first work by the COOL-LAMPS collaboration, we characterise the lensed galaxy using $griz$JH band photometry and optical and NIR spectroscopy. Using \code{Prospector}, we perform a stellar population synthesis analysis spectro-photometrically, to constrain the stellar mass, SFH, stellar metallicity and dust properties. Our dPIE-halo-based lens mass modeling from \code{LENSTOOL}, based on ground-based imaging, implies a median source magnification of at least $\sim$30, which puts the stellar mass and star formation rate (in the youngest age bin, closest to the epoch of observation) at logM$_{*}$ = $10.11^{+0.21}_{-0.26}$ and SFR = $27^{+13}_{-9}$ M$_{\odot}$/yr, respectively. This places this galaxy on the star-forming main sequence, at the massive end of that relation, and we find that \targ is 2-4x more luminous than a galaxy with the characteristic UV luminosity at these redshifts. We constrain a star formation history for \targ starting at $z\sim20$ that is consistent with constant star formation across $\sim$1 Gyr of cosmic time.

Current data are insufficient to constrain the metallicity and star formation history robustly. Also, as is almost always true, the lens model built on ground-based data limits the precision and accuracy of some aspects of our analysis; a lens model built from much sharper imaging (e.g., with HST) will provide insights into fluxes, flux variations, and identification of a variety of faint spatial features within this galaxy much more comprehensively via a proper source plane reconstruction. Moreover, our current spectroscopy only samples the apparently brightest clump in this galaxy. Once lensed into the observed giant arc, \targ is the brightest galaxy at $z>5$ currently known, and thus presents a unique opportunity for the detailed study of a massive galaxy immediately after the epoch of reionization. \targ is bright enough that precise high resolution observations will be possible across a broad range of wavelengths.

\acknowledgments

We would like to express gratitude towards the staff and workers at the 6.5m Magellan Telescopes at the Las Campanas Observatory, Chile, and the Gemini-North telescopes at Mauna Kea, Hawaii, USA, for their valuable labor. The authors wish to recognize and acknowledge the very significant cultural role and reverence that the summit of Mauna Kea has always had within the indigenous Hawaiian community. The authors are most fortunate to have the opportunity to conduct observations via the NOIRLab-operated Gemini-North facility from this mountain. \cref{GK acknowledges the impact of astrophysical observatories and research on this sacred Native Hawaiian land, and directs the reader towards resources on this matter for further consumption} (e.g., \citealp{2020arXiv200100970K,2020arXiv200100674P}, and references therein).

\cref{We would like to thank the anonymous reviewer for their thoughtful and helpful comments, that improved the quality of this manuscript and the robustness of ther results herein, substantially.} 

\cref{GK thanks Joel Leja and Benjamin Johnson for their responses, guidance and input related to specifics of the} \code{Prospector} framework. 

This work is supported by The College Undergraduate program at the University of Chicago, and the Department of Astronomy and Astrophysics at the University of Chicago. Magellan/PISCO operations are supported by NSF AST-1814719. This research was supported by an appointment to the NASA Postdoctoral Program at the NASA Goddard Space Flight Center administered by Universities Space Research Association through a contract with NASA. Work at Argonne National Lab is supported by UChicago Argonne LLC,Operator of Argonne National Laboratory (Argonne). Argonne, a U.S. Department of Energy Office of Science Laboratory, is operated under contract no. DE-AC02-06CH11357.

The Legacy Surveys consist of three individual and complementary projects: the Dark Energy Camera Legacy Survey (DECaLS; Proposal ID \#2014B-0404; PIs: David Schlegel and Arjun Dey), the Beijing-Arizona Sky Survey (BASS; NOAO Prop. ID \#2015A-0801; PIs: Zhou Xu and Xiaohui Fan), and the Mayall z-band Legacy Survey (MzLS; Prop. ID \#2016A-0453; PI: Arjun Dey). DECaLS, BASS and MzLS together include data obtained, respectively, at the Blanco telescope, Cerro Tololo Inter-American Observatory, NSF’s NOIRLab; the Bok telescope, Steward Observatory, University of Arizona; and the Mayall telescope, Kitt Peak National Observatory, NOIRLab. The Legacy Surveys project is honored to be permitted to conduct astronomical research on Iolkam Du’ag (Kitt Peak), a mountain with particular significance to the Tohono O’odham Nation.

NOIRLab is operated by the Association of Universities for Research in Astronomy (AURA) under a cooperative agreement with the National Science Foundation.

This project used data obtained with the Dark Energy Camera (DECam), which was constructed by the Dark Energy Survey (DES) collaboration. Funding for the DES Projects has been provided 19 by the U.S. Department of Energy, the U.S. National Science Foundation, the Ministry of Science and Education of Spain, the Science and Technology Facilities Council of the United Kingdom, the Higher Education Funding Council for England, the National Center for Supercomputing Applications at the University of Illinois at Urbana-Champaign, the Kavli Institute of Cosmological Physics at the University of Chicago, Center for Cosmology and Astro-Particle Physics at the Ohio State University, the Mitchell Institute for Fundamental Physics and Astronomy at Texas A\&M University, Financiadora de Estudos e Projetos, Fundacao Carlos Chagas Filho de Amparo, Financiadora de Estudos e Projetos, Fundacao Carlos Chagas Filho de Amparo a Pesquisa do Estado do Rio de Janeiro, Conselho Nacional de Desenvolvimento Cientifico e Tecnologico and the Ministerio da Ciencia, Tecnologia e Inovacao, the Deutsche Forschungsgemeinschaft and the Collaborating Institutions in the Dark Energy Survey. The Collaborating Institutions are Argonne National Laboratory, the University of California at Santa Cruz, the University of Cambridge, Centro de Investigaciones Energeticas, Medioambientales y Tecnologicas-Madrid, the University of Chicago, University College London, the DES-Brazil Consortium, the University of Edinburgh, the Eidgenossische Technische Hochschule (ETH) Zurich, Fermi National Accelerator Laboratory, the University of Illinois at Urbana-Champaign, the Institut de Ciencies de l’Espai (IEEC/CSIC), the Institut de Fisica d’Altes Energies, Lawrence Berkeley National Laboratory, the Ludwig Maximilians Universitat Munchen and the associated Excellence Cluster Universe, the University of Michigan, NSF’s NOIRLab, the University of Nottingham, the Ohio State University, the University of Pennsylvania, the University of Portsmouth, SLAC National Accelerator Laboratory, Stanford University, the University of Sussex, and Texas A\&M University.

The Legacy Surveys imaging of the DESI footprint is supported by the Director, Office of Science, Office of High Energy Physics of the U.S. Department of Energy under Contract No. DE-AC02- 05CH1123, by the National Energy Research Scientific Computing Center, a DOE Office of Science User Facility under the same contract; and by the U.S. National Science Foundation, Division of Astronomical Sciences under Contract No. AST-0950945 to NOIRLab.

\vspace{5mm}

\facilities{CTIO/4m Blanco Telescope, Magellan Telescopes 6.5m (Clay/LDSS3C, Clay/PISCO, Baade/FOURSTAR), Gemini-North Telescope (GNIRS)}

\software{\code{Python 3.6 — Prospector, python-FSPS, SEDpy, pygtc, Matplotlib, Numpy, Scipy, Astropy, LENSTOOL, Jupyter, IPython} Notebooks, GALFIT, SExtractor \citep{Bertin_1996}, SAO Image DS9, IRAF, IDL}

\bibliographystyle{yahapj}
\bibliography{main}

\begin{thebibliography}{}
\providecommand\natexlab[1]{#1}
\providecommand\JournalTitle[1]{#1}

\bibitem[{{Ahumada} {et~al.}(2020){Ahumada}, {Allende Prieto}, {Almeida},
  {Anders}, {Anderson}, {Andrews}, {Anguiano}, {Arcodia}, {Armengaud},
  {Aubert}, {Avila}, {Avila-Reese}, {Badenes}, {Balland }, {Barger},
  {Barrera-Ballesteros}, {Basu}, {Bautista}, {Beaton}, {Beers}, {Benavides},
  {Bender}, {Bernardi}, {Bershady}, {Beutler}, {Bidin}, {Bird}, {Bizyaev},
  {Blanc}, {Blanton}, {Boquien}, {Borissova}, {Bovy}, {Brand t}, {Brinkmann},
  {Brownstein}, {Bundy}, {Bureau}, {Burgasser}, {Burtin}, {Cano-D{\'\i}az},
  {Capasso}, {Cappellari}, {Carrera}, {Chabanier}, {Chaplin}, {Chapman},
  {Cherinka}, {Chiappini}, {Doohyun Choi}, {Chojnowski}, {Chung}, {Clerc},
  {Coffey}, {Comerford}, {Comparat}, {da Costa}, {Cousinou}, {Covey}, {Crane},
  {Cunha}, {da Silva Ilha}, {Dai}, {Damsted}, {Darling}, {Davidson}, {Davies},
  {Dawson}, {De}, {de la Macorra}, {De Lee}, {de Andrade Queiroz}, {Deconto
  Machado}, {de la Torre}, {Dell'Agli}, {du Mas des Bourboux},
  {Diamond-Stanic}, {Dillon}, {Donor}, {Drory}, {Duckworth}, {Dwelly},
  {Ebelke}, {Eftekharzadeh}, {Eigenbrot}, {Elsworth}, {Eracleous},
  {Erfanianfar}, {Escoffier}, {Fan}, {Farr}, {Fern{\'a}ndez-Trincado},
  {Feuillet}, {Finoguenov}, {Fofie}, {Fraser-McKelvie}, {Frinchaboy},
  {Fromenteau}, {Fu}, {Galbany}, {Garcia}, {Garc{\'\i}a-Hern{\'a}ndez}, {Garma
  Oehmichen}, {Ge}, {Geimba Maia}, {Geisler}, {Gelfand }, {Goddy},
  {Gonzalez-Perez}, {Grabowski}, {Green}, {Grier}, {Guo}, {Guy}, {Harding},
  {Hasselquist}, {Hawken}, {Hayes}, {Hearty}, {Hekker}, {Hogg}, {Holtzman},
  {Horta}, {Hou}, {Hsieh}, {Huber}, {Hunt}, {Ider Chitham}, {Imig}, {Jaber},
  {Jimenez Angel}, {Johnson}, {Jones}, {J{\"o}nsson}, {Jullo}, {Kim},
  {Kinemuchi}, {Kirkpatrick}, {Kite}, {Klaene}, {Kneib}, {Kollmeier}, {Kong},
  {Kounkel}, {Krishnarao}, {Lacerna}, {Lan}, {Lane}, {Law}, {Le Goff}, {Leung},
  {Lewis}, {Li}, {Lian}, {Lin}, {Long}, {Longa-Pe{\~n}a}, {Lundgren}, {Lyke},
  {Ted Mackereth}, {MacLeod}, {Majewski}, {Manchado}, {Maraston}, {Martini},
  {Masseron}, {Masters}, {Mathur}, {McDermid}, {Merloni}, {Merrifield},
  {M{\'e}sz{\'a}ros}, {Miglio}, {Minniti}, {Minsley}, {Miyaji}, {Mohammad},
  {Mosser}, {Mueller}, {Muna}, {Mu{\~n}oz-Guti{\'e}rrez}, {Myers}, {Nadathur},
  {Nair}, {Nandra}, {do Nascimento}, {Nevin}, {Newman}, {Nidever}, {Nitschelm},
  {Noterdaeme}, {O'Connell}, {Olmstead}, {Oravetz}, {Oravetz}, {Osorio},
  {Pace}, {Padilla}, {Palanque-Delabrouille}, {Palicio}, {Pan}, {Pan},
  {Parker}, {Paviot}, {Peirani}, {Pe{\~n}a Ram{\'r}ez}, {Penny}, {Percival},
  {Perez-Fournon}, {P{\'e}rez-R{\`a}fols}, {Petitjean}, {Pieri},
  {Pinsonneault}, {Poovelil}, {Povick}, {Prakash}, {Price-Whelan}, {Raddick},
  {Raichoor}, {Ray}, {Rembold}, {Rezaie}, {Riffel}, {Riffel}, {Rix}, {Robin},
  {Roman-Lopes}, {Rom{\'a}n-Z{\'u}{\~n}iga}, {Rose}, {Ross}, {Rossi}, {Rowland
  s}, {Rubin}, {Salvato}, {S{\'a}nchez}, {S{\'a}nchez-Menguiano},
  {S{\'a}nchez-Gallego}, {Sayres}, {Schaefer}, {Schiavon}, {Schimoia},
  {Schlafly}, {Schlegel}, {Schneider}, {Schultheis}, {Schwope}, {Seo},
  {Serenelli}, {Shafieloo}, {Shamsi}, {Shao}, {Shen}, {Shetrone}, {Shirley},
  {Silva Aguirre}, {Simon}, {Skrutskie}, {Slosar}, {Smethurst}, {Sobeck},
  {Sodi}, {Souto}, {Stark}, {Stassun}, {Steinmetz}, {Stello}, {Stermer},
  {Storchi-Bergmann}, {Streblyanska}, {Stringfellow}, {Stutz}, {Su{\'a}rez},
  {Sun}, {Taghizadeh-Popp}, {Talbot}, {Tayar}, {Thakar}, {Theriault}, {Thomas},
  {Thomas}, {Tinker}, {Tojeiro}, {Toledo}, {Tremonti}, {Troup}, {Tuttle},
  {Unda-Sanzana}, {Valentini}, {Vargas-Gonz{\'a}lez}, {Vargas-Maga{\~n}a},
  {V{\'a}zquez-Mata}, {Vivek}, {Wake}, {Wang}, {Weaver}, {Weijmans}, {Wild},
  {Wilson}, {Wilson}, {Wolthuis}, {Wood-Vasey}, {Yan}, {Yang}, {Y{\`e}che},
  {Zamora}, {Zarrouk}, {Zasowski}, {Zhang}, {Zhao}, {Zhao}, {Zheng}, {Zheng},
  {Zhu}, \& {Zou}}]{Ahumada_2020}
{Ahumada}, R., {Allende Prieto}, C., {Almeida}, A., {et~al.} 2020,
  \href{http://dx.doi.org/10.3847/1538-4365/ab929e}{\JournalTitle{\apjs}, 249,
  3}

\bibitem[{{Allam} {et~al.}(2007){Allam}, {Tucker}, {Lin}, {Diehl}, {Annis},
  {Buckley-Geer}, \& {Frieman}}]{Allam_2007}
{Allam}, S.~S., {Tucker}, D.~L., {Lin}, H., {et~al.} 2007,
  \href{http://dx.doi.org/10.1086/519520}{\JournalTitle{\apjl}, 662, L51}

\bibitem[{{Bayliss} {et~al.}(2011{\natexlab{a}}){Bayliss}, {Hennawi},
  {Gladders}, {Koester}, {Sharon}, {Dahle}, \& {Oguri}}]{Bayliss_2011}
{Bayliss}, M.~B., {Hennawi}, J.~F., {Gladders}, M.~D., {et~al.}
  2011{\natexlab{a}},
  \href{http://dx.doi.org/10.1088/0067-0049/193/1/8}{\JournalTitle{\apjs}, 193,
  8}

\bibitem[{{Bayliss} {et~al.}(2011{\natexlab{b}}){Bayliss}, {Hennawi},
  {Gladders}, {Koester}, {Sharon}, {Dahle}, \& {Oguri}}]{2011ApJS..193....8B}
---. 2011{\natexlab{b}},
  \href{http://dx.doi.org/10.1088/0067-0049/193/1/8}{\JournalTitle{\apjs}, 193,
  8}

\bibitem[{{Bayliss} {et~al.}(2014){Bayliss}, {Rigby}, {Sharon}, {Wuyts},
  {Florian}, {Gladders}, {Johnson}, \& {Oguri}}]{Bayliss_2014}
{Bayliss}, M.~B., {Rigby}, J.~R., {Sharon}, K., {et~al.} 2014,
  \href{http://dx.doi.org/10.1088/0004-637X/790/2/144}{\JournalTitle{\apj},
  790, 144}

\bibitem[{{Belokurov} {et~al.}(2007){Belokurov}, {Evans}, {Moiseev}, {King},
  {Hewett}, {Pettini}, {Wyrzykowski}, {McMahon}, {Smith}, {Gilmore}, {Sanchez},
  {Udalski}, {Koposov}, {Zucker}, \& {Walcher}}]{Belokurov_2007}
{Belokurov}, V., {Evans}, N.~W., {Moiseev}, A., {et~al.} 2007,
  \href{http://dx.doi.org/10.1086/524948}{\JournalTitle{\apjl}, 671, L9}

\bibitem[{{Bertin} \& {Arnouts}(1996)}]{Bertin_1996}
{Bertin}, E., \& {Arnouts}, S. 1996,
  \href{http://dx.doi.org/10.1051/aas:1996164}{\JournalTitle{\aaps}, 117, 393}

\bibitem[{{Bicker} \& {Fritze-v. Alvensleben}(2005)}]{Bicker_2005}
{Bicker}, J., \& {Fritze-v. Alvensleben}, U. 2005,
  \href{http://dx.doi.org/10.1051/0004-6361:200500194}{\JournalTitle{\aap},
  443, L19}

\bibitem[{{Bleem} {et~al.}(2015){Bleem}, {Stalder}, {de Haan}, {Aird}, {Allen},
  {Applegate}, {Ashby}, {Bautz}, {Bayliss}, {Benson}, {Bocquet}, {Brodwin},
  {Carlstrom}, {Chang}, {Chiu}, {Cho}, {Clocchiatti}, {Crawford}, {Crites},
  {Desai}, {Dietrich}, {Dobbs}, {Foley}, {Forman}, {George}, {Gladders},
  {Gonzalez}, {Halverson}, {Hennig}, {Hoekstra}, {Holder}, {Holzapfel},
  {Hrubes}, {Jones}, {Keisler}, {Knox}, {Lee}, {Leitch}, {Liu}, {Lueker},
  {Luong-Van}, {Mantz}, {Marrone}, {McDonald}, {McMahon}, {Meyer}, {Mocanu},
  {Mohr}, {Murray}, {Padin}, {Pryke}, {Reichardt}, {Rest}, {Ruel}, {Ruhl},
  {Saliwanchik}, {Saro}, {Sayre}, {Schaffer}, {Schrabback}, {Shirokoff},
  {Song}, {Spieler}, {Stanford}, {Staniszewski}, {Stark}, {Story}, {Stubbs},
  {Vand erlinde}, {Vieira}, {Vikhlinin}, {Williamson}, {Zahn}, \&
  {Zenteno}}]{Bleem_2015}
{Bleem}, L.~E., {Stalder}, B., {de Haan}, T., {et~al.} 2015,
  \href{http://dx.doi.org/10.1088/0067-0049/216/2/27}{\JournalTitle{\apjs},
  216, 27}

\bibitem[{{Bleem} {et~al.}(2020){Bleem}, {Bocquet}, {Stalder}, {Gladders},
  {Ade}, {Allen}, {Anderson}, {Annis}, {Ashby}, {Austermann}, {Avila}, {Avva},
  {Bayliss}, {Beall}, {Bechtol}, {Bender}, {Benson}, {Bertin}, {Bianchini},
  {Blake}, {Brodwin}, {Brooks}, {Buckley-Geer}, {Burke}, {Carlstrom}, {Rosell},
  {Carrasco Kind}, {Carretero}, {Chang}, {Chiang}, {Citron}, {Moran},
  {Costanzi}, {Crawford}, {Crites}, {da Costa}, {de Haan}, {De Vicente},
  {Desai}, {Diehl}, {Dietrich}, {Dobbs}, {Eifler}, {Everett}, {Flaugher},
  {Floyd}, {Frieman}, {Gallicchio}, {Garc{\'\i}a-Bellido}, {George}, {Gerdes},
  {Gilbert}, {Gruen}, {Gruendl}, {Gschwend}, {Gupta}, {Gutierrez}, {Halverson},
  {Harrington}, {Henning}, {Heymans}, {Holder}, {Hollowood}, {Holzapfel},
  {Honscheid}, {Hrubes}, {Huang}, {Hubmayr}, {Irwin}, {James}, {Jeltema},
  {Joudaki}, {Khullar}, {Klein}, {Knox}, {Kuropatkin}, {Lee}, {Li}, {Lidman},
  {Lowitz}, {MacCrann}, {Mahler}, {Maia}, {Marshall}, {McDonald}, {McMahon},
  {Melchior}, {Menanteau}, {Meyer}, {Miquel}, {Mocanu}, {Mohr}, {Montgomery},
  {Nadolski}, {Natoli}, {Nibarger}, {Noble}, {Novosad}, {Padin}, {Palmese},
  {Parkinson}, {Patil}, {Paz-Chinch{\'o}n}, {Plazas}, {Pryke}, {Ramachandra},
  {Reichardt}, {Remolina Gonz{\'a}lez}, {Romer}, {Roodman}, {Ruhl}, {Rykoff},
  {Saliwanchik}, {Sanchez}, {Saro}, {Sayre}, {Schaffer}, {Schrabback},
  {Serrano}, {Sharon}, {Sievers}, {Smecher}, {Smith}, {Soares-Santos}, {Stark},
  {Story}, {Suchyta}, {Tarle}, {Tucker}, {Vanderlinde}, {Veach}, {Vieira},
  {Wang}, {Weller}, {Whitehorn}, {Wu}, {Yefremenko}, \& {Zhang}}]{Bleem_2020}
{Bleem}, L.~E., {Bocquet}, S., {Stalder}, B., {et~al.} 2020,
  \href{http://dx.doi.org/10.3847/1538-4365/ab6993}{\JournalTitle{\apjs}, 247,
  25}

\bibitem[{Bocquet \& Carter(2016)}]{Bocquet2016}
Bocquet, S., \& Carter, F.~W. 2016,
  \href{http://dx.doi.org/10.21105/joss.00046}{\JournalTitle{The Journal of
  Open Source Software}, 1}

\bibitem[{{Bouwens} {et~al.}(2014){Bouwens}, {Illingworth}, {Oesch},
  {Labb{\'e}}, {van Dokkum}, {Trenti}, {Franx}, {Smit}, {Gonzalez}, \&
  {Magee}}]{Bouwens_2014}
{Bouwens}, R.~J., {Illingworth}, G.~D., {Oesch}, P.~A., {et~al.} 2014,
  \href{http://dx.doi.org/10.1088/0004-637X/793/2/115}{\JournalTitle{\apj},
  793, 115}

\bibitem[{{Bouwens} {et~al.}(2015){Bouwens}, {Illingworth}, {Oesch}, {Trenti},
  {Labb{\'e}}, {Bradley}, {Carollo}, {van Dokkum}, {Gonzalez}, {Holwerda},
  {Franx}, {Spitler}, {Smit}, \& {Magee}}]{Bouwens_2015}
---. 2015,
  \href{http://dx.doi.org/10.1088/0004-637X/803/1/34}{\JournalTitle{\apj}, 803,
  34}

\bibitem[{{Calzetti} {et~al.}(2000){Calzetti}, {Armus}, {Bohlin}, {Kinney},
  {Koornneef}, \& {Storchi-Bergmann}}]{Calzetti_2000}
{Calzetti}, D., {Armus}, L., {Bohlin}, R.~C., {et~al.} 2000,
  \href{http://dx.doi.org/10.1086/308692}{\JournalTitle{\apj}, 533, 682}

\bibitem[{{Calzetti} {et~al.}(1994){Calzetti}, {Kinney}, \&
  {Storchi-Bergmann}}]{Calzetti_1994}
{Calzetti}, D., {Kinney}, A.~L., \& {Storchi-Bergmann}, T. 1994,
  \href{http://dx.doi.org/10.1086/174346}{\JournalTitle{\apj}, 429, 582}

\bibitem[{{Casey} {et~al.}(2019){Casey}, {Zavala}, {Aravena}, {B{\'e}thermin},
  {Caputi}, {Champagne}, {Clements}, {da Cunha}, {Drew}, {Finkelstein},
  {Hayward}, {Kartaltepe}, {Knudsen}, {Koekemoer}, {Magdis}, {Man}, {Manning},
  {Scoville}, {Sheth}, {Spilker}, {Staguhn}, {Talia}, {Taniguchi}, {Toft},
  {Treister}, \& {Yun}}]{Casey_2019}
{Casey}, C.~M., {Zavala}, J.~A., {Aravena}, M., {et~al.} 2019,
  \href{http://dx.doi.org/10.3847/1538-4357/ab52ff}{\JournalTitle{\apj}, 887,
  55}

\bibitem[{Chambers {et~al.}(2016)Chambers, Magnier, Metcalfe, Flewelling,
  Huber, Waters, Denneau, Draper, Farrow, Finkbeiner, Holmberg, Koppenhoefer,
  Price, Rest, Saglia, Schlafly, Smartt, Sweeney, Wainscoat, Burgett, Chastel,
  Grav, Heasley, Hodapp, Jedicke, Kaiser, Kudritzki, Luppino, Lupton, Monet,
  Morgan, Onaka, Shiao, Stubbs, Tonry, White, Bañados, Bell, Bender, Bernard,
  Boegner, Boffi, Botticella, Calamida, Casertano, Chen, Chen, Cole, Deacon,
  Frenk, Fitzsimmons, Gezari, Gibbs, Goessl, Goggia, Gourgue, Goldman, Grant,
  Grebel, Hambly, Hasinger, Heavens, Heckman, Henderson, Henning, Holman, Hopp,
  Ip, Isani, Jackson, Keyes, Koekemoer, Kotak, Le, Liska, Long, Lucey, Liu,
  Martin, Masci, McLean, Mindel, Misra, Morganson, Murphy, Obaika, Narayan,
  Nieto-Santisteban, Norberg, Peacock, Pier, Postman, Primak, Rae, Rai, Riess,
  Riffeser, Rix, Röser, Russel, Rutz, Schilbach, Schultz, Scolnic, Strolger,
  Szalay, Seitz, Small, Smith, Soderblom, Taylor, Thomson, Taylor, Thakar,
  Thiel, Thilker, Unger, Urata, Valenti, Wagner, Walder, Walter, Watters,
  Werner, Wood-Vasey, \& Wyse}]{chambers2016panstarrs1}
Chambers, K.~C., Magnier, E.~A., Metcalfe, N., {et~al.} 2016, The Pan-STARRS1
  Surveys, \href{http://arxiv.org/abs/1612.05560}{{\sffamily arXiv:1612.05560
  [astro-ph.IM]}}

\bibitem[{{Choi} {et~al.}(2016){Choi}, {Dotter}, {Conroy}, {Cantiello},
  {Paxton}, \& {Johnson}}]{Choi_2016}
{Choi}, J., {Dotter}, A., {Conroy}, C., {et~al.} 2016,
  \href{http://dx.doi.org/10.3847/0004-637X/823/2/102}{\JournalTitle{\apj},
  823, 102}

\bibitem[{{Coe} {et~al.}(2019){Coe}, {Salmon}, {Brada{\v{c}}}, {Bradley},
  {Sharon}, {Zitrin}, {Acebron}, {Cerny}, {Cibirka}, {Strait},
  {Paterno-Mahler}, {Mahler}, {Avila}, {Ogaz}, {Huang}, {Pelliccia}, {Stark},
  {Mainali}, {Oesch}, {Trenti}, {Carrasco}, {Dawson}, {Rodney}, {Strolger},
  {Riess}, {Jones}, {Frye}, {Czakon}, {Umetsu}, {Vulcani}, {Graur}, {Jha},
  {Graham}, {Molino}, {Nonino}, {Hjorth}, {Selsing}, {Christensen},
  {Kikuchihara}, {Ouchi}, {Oguri}, {Welch}, {Lemaux}, {Andrade-Santos}, {Hoag},
  {Johnson}, {Peterson}, {Past}, {Fox}, {Agulli}, {Livermore}, {Ryan}, {Lam},
  {Sendra-Server}, {Toft}, {Lovisari}, \& {Su}}]{Coe_2019}
{Coe}, D., {Salmon}, B., {Brada{\v{c}}}, M., {et~al.} 2019,
  \href{http://dx.doi.org/10.3847/1538-4357/ab412b}{\JournalTitle{\apj}, 884,
  85}

\bibitem[{{Conroy} \& {Gunn}(2010)}]{Conroy_2010}
{Conroy}, C., \& {Gunn}, J.~E. 2010,
  \href{http://dx.doi.org/10.1088/0004-637X/712/2/833}{\JournalTitle{\apj},
  712, 833}

\bibitem[{{Cornachione} {et~al.}(2018){Cornachione}, {Bolton}, {Shu}, {Zheng},
  {Montero-Dorta}, {Brownstein}, {Oguri}, {Kochanek}, {Mao},
  {P{\`e}rez-Fournon}, {Marques-Chaves}, \& {M{\`e}nard}}]{Cornachione_2018}
{Cornachione}, M.~A., {Bolton}, A.~S., {Shu}, Y., {et~al.} 2018,
  \href{http://dx.doi.org/10.3847/1538-4357/aaa412}{\JournalTitle{\apj}, 853,
  148}

\bibitem[{{de Vaucouleurs}(1948)}]{1948AnAp...11..247D}
{de Vaucouleurs}, G. 1948, \JournalTitle{Annales d'Astrophysique}, 11, 247

\bibitem[{{Dey} {et~al.}(2019){Dey}, {Schlegel}, {Lang}, {Blum}, {Burleigh},
  {Fan}, {Findlay}, {Finkbeiner}, {Herrera}, {Juneau}, {Landriau}, {Levi},
  {McGreer}, {Meisner}, {Myers}, {Moustakas}, {Nugent}, {Patej}, {Schlafly},
  {Walker}, {Valdes}, {Weaver}, {Y{\`e}che}, {Zou}, {Zhou}, {Abareshi},
  {Abbott}, {Abolfathi}, {Aguilera}, {Alam}, {Allen}, {Alvarez}, {Annis},
  {Ansarinejad}, {Aubert}, {Beechert}, {Bell}, {BenZvi}, {Beutler}, {Bielby},
  {Bolton}, {Brice{\~n}o}, {Buckley-Geer}, {Butler}, {Calamida}, {Carlberg},
  {Carter}, {Casas}, {Castander}, {Choi}, {Comparat}, {Cukanovaite}, {Delubac},
  {DeVries}, {Dey}, {Dhungana}, {Dickinson}, {Ding}, {Donaldson}, {Duan},
  {Duckworth}, {Eftekharzadeh}, {Eisenstein}, {Etourneau}, {Fagrelius},
  {Farihi}, {Fitzpatrick}, {Font-Ribera}, {Fulmer}, {G{\"a}nsicke},
  {Gaztanaga}, {George}, {Gerdes}, {Gontcho}, {Gorgoni}, {Green}, {Guy},
  {Harmer}, {Hernand ez}, {Honscheid}, {Huang}, {James}, {Jannuzi}, {Jiang},
  {Joyce}, {Karcher}, {Karkar}, {Kehoe}, {Kneib}, {Kueter-Young}, {Lan},
  {Lauer}, {Le Guillou}, {Le Van Suu}, {Lee}, {Lesser}, {Perreault Levasseur},
  {Li}, {Mann}, {Marshall}, {Mart{\'\i}nez-V{\'a}zquez}, {Martini}, {du Mas des
  Bourboux}, {McManus}, {Meier}, {M{\'e}nard}, {Metcalfe},
  {Mu{\~n}oz-Guti{\'e}rrez}, {Najita}, {Napier}, {Narayan}, {Newman}, {Nie},
  {Nord}, {Norman}, {Olsen}, {Paat}, {Palanque-Delabrouille}, {Peng},
  {Poppett}, {Poremba}, {Prakash}, {Rabinowitz}, {Raichoor}, {Rezaie},
  {Robertson}, {Roe}, {Ross}, {Ross}, {Rudnick}, {Safonova}, {Saha},
  {S{\'a}nchez}, {Savary}, {Schweiker}, {Scott}, {Seo}, {Shan}, {Silva},
  {Slepian}, {Soto}, {Sprayberry}, {Staten}, {Stillman}, {Stupak}, {Summers},
  {Sien Tie}, {Tirado}, {Vargas-Maga{\~n}a}, {Vivas}, {Wechsler}, {Williams},
  {Yang}, {Yang}, {Yapici}, {Zaritsky}, {Zenteno}, {Zhang}, {Zhang}, {Zhou}, \&
  {Zhou}}]{Dey_2019}
{Dey}, A., {Schlegel}, D.~J., {Lang}, D., {et~al.} 2019,
  \href{http://dx.doi.org/10.3847/1538-3881/ab089d}{\JournalTitle{\aj}, 157,
  168}

\bibitem[{Diehl {et~al.}(2017)Diehl, Buckley-Geer, Lindgren, Nord, Gaitsch,
  Gaitsch, Lin, Allam, Collett, Furlanetto, Gill, More, Nightingale, Odden,
  Pellico, Tucker, da~Costa, Neto, Kuropatkin, Soares-Santos, Welch, Zhang,
  Frieman, Abdalla, Annis, Benoit-L{\'{e}}vy, Bertin, Brooks, Burke, Rosell,
  Kind, Carretero, Cunha, D'Andrea, Desai, Dietrich, Drlica-Wagner, Evrard,
  Finley, Flaugher, Garc{\'{\i}}a-Bellido, Gerdes, Goldstein, Gruen, Gruendl,
  Gschwend, Gutierrez, James, Kuehn, Kuhlmann, Lahav, Li, Lima, Maia, Marshall,
  Menanteau, Miquel, Nichol, Nugent, Ogando, Plazas, Reil, Romer, Sako,
  Sanchez, Santiago, Scarpine, Schindler, Schubnell, Sevilla-Noarbe, Sheldon,
  Smith, Sobreira, Suchyta, Swanson, Tarle, Thomas, \& and}]{Diehl_2017}
Diehl, H.~T., Buckley-Geer, E.~J., Lindgren, K.~A., {et~al.} 2017,
  \href{http://dx.doi.org/10.3847/1538-4365/aa8667}{\JournalTitle{The
  Astrophysical Journal Supplement Series}, 232, 15}

\bibitem[{{Elias} {et~al.}(2006){Elias}, {Joyce}, {Liang}, {Muller}, {Hileman},
  \& {George}}]{Elias_2006}
{Elias}, J.~H., {Joyce}, R.~R., {Liang}, M., {et~al.} 2006,
  \href{http://dx.doi.org/10.1117/12.671817}{in Society of Photo-Optical
  Instrumentation Engineers (SPIE) Conference Series, Vol. 6269, Society of
  Photo-Optical Instrumentation Engineers (SPIE) Conference Series}, 62694C

\bibitem[{{El{\'\i}asd{\'o}ttir} {et~al.}(2007){El{\'\i}asd{\'o}ttir},
  {Limousin}, {Richard}, {Hjorth}, {Kneib}, {Natarajan}, {Pedersen}, {Jullo},
  \& {Paraficz}}]{2007arXiv0710.5636E}
{El{\'\i}asd{\'o}ttir}, {\'A}., {Limousin}, M., {Richard}, J., {et~al.} 2007,
  \JournalTitle{arXiv e-prints}, arXiv:0710.5636

\bibitem[{{Falc{\'o}n-Barroso} {et~al.}(2011){Falc{\'o}n-Barroso},
  {S{\'a}nchez-Bl{\'a}zquez}, {Vazdekis}, {Ricciardelli}, {Cardiel}, {Cenarro},
  {Gorgas}, \& {Peletier}}]{2011A&A...532A..95F}
{Falc{\'o}n-Barroso}, J., {S{\'a}nchez-Bl{\'a}zquez}, P., {Vazdekis}, A.,
  {et~al.} 2011,
  \href{http://dx.doi.org/10.1051/0004-6361/201116842}{\JournalTitle{\aap},
  532, A95}

\bibitem[{{Finkelstein} {et~al.}(2012){Finkelstein}, {Papovich}, {Salmon},
  {Finlator}, {Dickinson}, {Ferguson}, {Giavalisco}, {Koekemoer}, {Reddy},
  {Bassett}, {Conselice}, {Dunlop}, {Faber}, {Grogin}, {Hathi}, {Kocevski},
  {Lai}, {Lee}, {McLure}, {Mobasher}, \& {Newman}}]{Finkelstein_2012}
{Finkelstein}, S.~L., {Papovich}, C., {Salmon}, B., {et~al.} 2012,
  \href{http://dx.doi.org/10.1088/0004-637X/756/2/164}{\JournalTitle{\apj},
  756, 164}

\bibitem[{{Florian} {et~al.}(2020){Florian}, {Rigby}, {Acharyya}, {Sharon},
  {Gladders}, {Kewley}, {Khullar}, {Gozman}, {Brammer}, {Momcheva}, {Nicholls},
  {LaMassa}, {Dahle}, {Bayliss}, {Wuyts}, {Johnson}, \&
  {Whitaker}}]{Florian_2020}
{Florian}, M.~K., {Rigby}, J.~R., {Acharyya}, A., {et~al.} 2020,
  \JournalTitle{arXiv e-prints}, arXiv:2006.11387

\bibitem[{{Foreman-Mackey} {et~al.}(2013){Foreman-Mackey}, {Hogg}, {Lang}, \&
  {Goodman}}]{2013PASP..125..306F}
{Foreman-Mackey}, D., {Hogg}, D.~W., {Lang}, D., \& {Goodman}, J. 2013,
  \href{http://dx.doi.org/10.1086/670067}{\JournalTitle{\pasp}, 125, 306}

\bibitem[{{Franx} {et~al.}(1997){Franx}, {Illingworth}, {Kelson}, {van Dokkum},
  \& {Tran}}]{Franx_1997}
{Franx}, M., {Illingworth}, G.~D., {Kelson}, D.~D., {van Dokkum}, P.~G., \&
  {Tran}, K.-V. 1997,
  \href{http://dx.doi.org/10.1086/310844}{\JournalTitle{\apjl}, 486, L75}

\bibitem[{{Frye} {et~al.}(2002){Frye}, {Broadhurst}, \&
  {Ben{\'\i}tez}}]{Frye_2002}
{Frye}, B., {Broadhurst}, T., \& {Ben{\'\i}tez}, N. 2002,
  \href{http://dx.doi.org/10.1086/338965}{\JournalTitle{\apj}, 568, 558}

\bibitem[{{Garg} {et~al.}(2007){Garg}, {Stubbs}, {Challis}, {Wood-Vasey},
  {Blondin}, {Huber}, {Cook}, {Nikolaev}, {Rest}, {Smith}, {Olsen}, {Suntzeff},
  {Aguilera}, {Prieto}, {Becker}, {Miceli}, {Miknaitis}, {Clocchiatti},
  {Minniti}, {Morelli}, \& {Welch}}]{Garg_2007}
{Garg}, A., {Stubbs}, C.~W., {Challis}, P., {et~al.} 2007,
  \href{http://dx.doi.org/10.1086/510118}{\JournalTitle{\aj}, 133, 403}

\bibitem[{{Gilbank} {et~al.}(2010){Gilbank}, {Baldry}, {Balogh}, {Glazebrook},
  \& {Bower}}]{Gilbank_2010}
{Gilbank}, D.~G., {Baldry}, I.~K., {Balogh}, M.~L., {Glazebrook}, K., \&
  {Bower}, R.~G. 2010,
  \href{http://dx.doi.org/10.1111/j.1365-2966.2010.16640.x}{\JournalTitle{\mnras},
  405, 2594}

\bibitem[{{Gladders} {et~al.}(2002){Gladders}, {Yee}, \&
  {Ellingson}}]{Gladders_2002}
{Gladders}, M.~D., {Yee}, H.~K.~C., \& {Ellingson}, E. 2002,
  \href{http://dx.doi.org/10.1086/324637}{\JournalTitle{\aj}, 123, 1}

\bibitem[{{Hayes} {et~al.}(2011){Hayes}, {Schaerer}, {{\"O}stlin}, {Mas-Hesse},
  {Atek}, \& {Kunth}}]{Hayes_2011}
{Hayes}, M., {Schaerer}, D., {{\"O}stlin}, G., {et~al.} 2011,
  \href{http://dx.doi.org/10.1088/0004-637X/730/1/8}{\JournalTitle{\apj}, 730,
  8}

\bibitem[{{Hinshaw} {et~al.}(2013){Hinshaw}, {Larson}, {Komatsu}, {Spergel},
  {Bennett}, {Dunkley}, {Nolta}, {Halpern}, {Hill}, {Odegard}, {Page}, {Smith},
  {Weiland}, {Gold}, {Jarosik}, {Kogut}, {Limon}, {Meyer}, {Tucker}, {Wollack},
  \& {Wright}}]{Hinshaw_2013}
{Hinshaw}, G., {Larson}, D., {Komatsu}, E., {et~al.} 2013,
  \href{http://dx.doi.org/10.1088/0067-0049/208/2/19}{\JournalTitle{\apjs},
  208, 19}

\bibitem[{{Huang} {et~al.}(2020){Huang}, {Storfer}, {Ravi}, {Pilon}, {Domingo},
  {Schlegel}, {Bailey}, {Dey}, {Gupta}, {Herrera}, {Juneau}, {Landriau},
  {Lang}, {Meisner}, {Moustakas}, {Myers}, {Schlafly}, {Valdes}, {Weaver},
  {Yang}, \& {Y{\`e}che}}]{Huang_2020}
{Huang}, X., {Storfer}, C., {Ravi}, V., {et~al.} 2020,
  \href{http://dx.doi.org/10.3847/1538-4357/ab7ffb}{\JournalTitle{\apj}, 894,
  78}

\bibitem[{{Inami} {et~al.}(2017){Inami}, {Bacon}, {Brinchmann}, {Richard},
  {Contini}, {Conseil}, {Hamer}, {Akhlaghi}, {Bouch{\'e}}, {Cl{\'e}ment},
  {Desprez}, {Drake}, {Hashimoto}, {Leclercq}, {Maseda}, {Michel-Dansac},
  {Paalvast}, {Tresse}, {Ventou}, {Kollatschny}, {Boogaard}, {Finley},
  {Marino}, {Schaye}, \& {Wisotzki}}]{Inami_2017}
{Inami}, H., {Bacon}, R., {Brinchmann}, J., {et~al.} 2017,
  \href{http://dx.doi.org/10.1051/0004-6361/201731195}{\JournalTitle{\aap},
  608, A2}

\bibitem[{Ivison {et~al.}(2020)Ivison, Richard, Biggs, Zwaan, Falgarone,
  Arumugam, van der Werf, \& Rujopakarn}]{Ivison_2020}
Ivison, R.~J., Richard, J., Biggs, A.~D., {et~al.} 2020,
  \href{http://dx.doi.org/10.1093/mnrasl/slaa046}{\JournalTitle{Monthly Notices
  of the Royal Astronomical Society: Letters}, 495, L1}

\bibitem[{Jacobs {et~al.}(2019)Jacobs, Collett, Glazebrook, Buckley-Geer,
  Diehl, Lin, McCarthy, Qin, Odden, Escudero, Dial, Yung, Gaitsch, Pellico,
  Lindgren, Abbott, Annis, Avila, Brooks, Burke, Rosell, Kind, Carretero,
  da~Costa, Vicente, Fosalba, Frieman, Garc{\'{\i}}a-Bellido, Gaztanaga,
  Goldstein, Gruen, Gruendl, Gschwend, Hollowood, Honscheid, Hoyle, James,
  Krause, Kuropatkin, Lahav, Lima, Maia, Marshall, Miquel, Plazas, Roodman,
  Sanchez, Scarpine, Serrano, Sevilla-Noarbe, Smith, Sobreira, Suchyta,
  Swanson, Tarle, Vikram, Walker, \& and}]{Jacobs_2019}
Jacobs, C., Collett, T., Glazebrook, K., {et~al.} 2019,
  \href{http://dx.doi.org/10.3847/1538-4365/ab26b6}{\JournalTitle{The
  Astrophysical Journal Supplement Series}, 243, 17}

\bibitem[{{Jaelani} {et~al.}(2020){Jaelani}, {More}, {Oguri}, {Sonnenfeld},
  {Suyu}, {Rusu}, {Wong}, {Chan}, {Kayo}, {Lee}, {Chao}, {Coupon}, {Inoue}, \&
  {Futamase}}]{Jaelani_2020}
{Jaelani}, A.~T., {More}, A., {Oguri}, M., {et~al.} 2020,
  \href{http://dx.doi.org/10.1093/mnras/staa1062}{\JournalTitle{\mnras}, 495,
  1291}

\bibitem[{{James} {et~al.}(2014){James}, {Pettini}, {Christensen}, {Auger},
  {Becker}, {King}, {Quider}, {Shapley}, \& {Steidel}}]{James_2014}
{James}, B.~L., {Pettini}, M., {Christensen}, L., {et~al.} 2014,
  \href{http://dx.doi.org/10.1093/mnras/stu287}{\JournalTitle{\mnras}, 440,
  1794}

\bibitem[{{Jarrett} {et~al.}(2000){Jarrett}, {Chester}, {Cutri}, {Schneider},
  {Skrutskie}, \& {Huchra}}]{Jarrett_2000}
{Jarrett}, T.~H., {Chester}, T., {Cutri}, R., {et~al.} 2000,
  \href{http://dx.doi.org/10.1086/301330}{\JournalTitle{\aj}, 119, 2498}

\bibitem[{Johnson \& Leja(2017)}]{prospector}
Johnson, B., \& Leja, J. 2017, bd-j/prospector: Initial release

\bibitem[{{Johnson} {et~al.}(2017){Johnson}, {Rigby}, {Sharon}, {Gladders},
  {Florian}, {Bayliss}, {Wuyts}, {Whitaker}, {Livermore}, \&
  {Murray}}]{Johnson_2017}
{Johnson}, T.~L., {Rigby}, J.~R., {Sharon}, K., {et~al.} 2017,
  \href{http://dx.doi.org/10.3847/2041-8213/aa7516}{\JournalTitle{\apjl}, 843,
  L21}

\bibitem[{Johnson {et~al.}(2017)Johnson, Sharon, Gladders, Rigby, Bayliss,
  Wuyts, Whitaker, Florian, \& Murray}]{Johnson_2017_2}
Johnson, T.~L., Sharon, K., Gladders, M.~D., {et~al.} 2017,
  \href{http://dx.doi.org/10.3847/1538-4357/aa7756}{\JournalTitle{The
  Astrophysical Journal}, 843, 78}

\bibitem[{{Jullo} {et~al.}(2007){Jullo}, {Kneib}, {Limousin},
  {El{\'\i}asd{\'o}ttir}, {Marshall}, \& {Verdugo}}]{Jullo_2007}
{Jullo}, E., {Kneib}, J.~P., {Limousin}, M., {et~al.} 2007,
  \href{http://dx.doi.org/10.1088/1367-2630/9/12/447}{\JournalTitle{New Journal
  of Physics}, 9, 447}

\bibitem[{{Kahanamoku} {et~al.}(2020){Kahanamoku}, {'Anolani Alegado},
  {Kagawa-Viviani}, {Leimomi Kamelamela}, {Kamai}, {Walkowicz},
  {Prescod-Weinstein}, {Alexa de los Reyes}, \&
  {Neilson}}]{2020arXiv200100970K}
{Kahanamoku}, S., {'Anolani Alegado}, R., {Kagawa-Viviani}, A., {et~al.} 2020,
  \JournalTitle{arXiv e-prints}, arXiv:2001.00970

\bibitem[{Kass \& Raftery(1995)}]{kass1995bayes}
Kass, R.~E., \& Raftery, A.~E. 1995, \JournalTitle{Journal of the american
  statistical association}, 90, 773

\bibitem[{{Kennicutt}(1992)}]{Kennicutt_1992}
{Kennicutt}, Robert~C., J. 1992,
  \href{http://dx.doi.org/10.1086/171154}{\JournalTitle{\apj}, 388, 310}

\bibitem[{{Kewley} {et~al.}(2004){Kewley}, {Geller}, \& {Jansen}}]{Kewley_2004}
{Kewley}, L.~J., {Geller}, M.~J., \& {Jansen}, R.~A. 2004,
  \href{http://dx.doi.org/10.1086/382723}{\JournalTitle{\aj}, 127, 2002}

\bibitem[{Kneib \& Soucail(1996)}]{Kneib_1996}
Kneib, J.~P., \& Soucail, G. 1996, in Astrophysical Applications of
  Gravitational Lensing, ed. C.~S. Kochanek \& J.~N. Hewitt (Dordrecht:
  Springer Netherlands), 113

\bibitem[{{Koester} {et~al.}(2010){Koester}, {Gladders}, {Hennawi}, {Sharon},
  {Wuyts}, {Rigby}, {Bayliss}, \& {Dahle}}]{Koester_2010}
{Koester}, B.~P., {Gladders}, M.~D., {Hennawi}, J.~F., {et~al.} 2010,
  \href{http://dx.doi.org/10.1088/2041-8205/723/1/L73}{\JournalTitle{\apjl},
  723, L73}

\bibitem[{{Kroupa}(2001)}]{Kroupa_2001}
{Kroupa}, P. 2001,
  \href{http://dx.doi.org/10.1046/j.1365-8711.2001.04022.x}{\JournalTitle{\mnras},
  322, 231}

\bibitem[{{Kubo} {et~al.}(2010){Kubo}, {Allam}, {Drabek}, {Lin}, {Tucker},
  {Buckley-Geer}, {Diehl}, {Soares-Santos}, {Hao}, {Wiesner}, {West}, {Kubik},
  {Annis}, \& {Frieman}}]{Kubo_2010}
{Kubo}, J.~M., {Allam}, S.~S., {Drabek}, E., {et~al.} 2010,
  \href{http://dx.doi.org/10.1088/2041-8205/724/2/L137}{\JournalTitle{\apjl},
  724, L137}

\bibitem[{{Leja} {et~al.}(2019){Leja}, {Carnall}, {Johnson}, {Conroy}, \&
  {Speagle}}]{Leja_2019}
{Leja}, J., {Carnall}, A.~C., {Johnson}, B.~D., {Conroy}, C., \& {Speagle},
  J.~S. 2019,
  \href{http://dx.doi.org/10.3847/1538-4357/ab133c}{\JournalTitle{\apj}, 876,
  3}

\bibitem[{{Leja} {et~al.}(2017){Leja}, {Johnson}, {Conroy}, {van Dokkum}, \&
  {Byler}}]{Leja_2017}
{Leja}, J., {Johnson}, B.~D., {Conroy}, C., {van Dokkum}, P.~G., \& {Byler}, N.
  2017, \href{http://dx.doi.org/10.3847/1538-4357/aa5ffe}{\JournalTitle{\apj},
  837, 170}

\bibitem[{{Licquia} \& {Newman}(2015)}]{2015ApJ...806...96L}
{Licquia}, T.~C., \& {Newman}, J.~A. 2015,
  \href{http://dx.doi.org/10.1088/0004-637X/806/1/96}{\JournalTitle{\apj}, 806,
  96}

\bibitem[{{Livermore} {et~al.}(2015){Livermore}, {Jones}, {Richard}, {Bower},
  {Swinbank}, {Yuan}, {Edge}, {Ellis}, {Kewley}, {Smail}, {Coppin}, \&
  {Ebeling}}]{Livermore_2015}
{Livermore}, R.~C., {Jones}, T.~A., {Richard}, J., {et~al.} 2015,
  \href{http://dx.doi.org/10.1093/mnras/stv686}{\JournalTitle{\mnras}, 450,
  1812}

\bibitem[{{Lotz} {et~al.}(2017){Lotz}, {Koekemoer}, {Coe}, {Grogin}, {Capak},
  {Mack}, {Anderson}, {Avila}, {Barker}, {Borncamp}, {Brammer}, {Durbin},
  {Gunning}, {Hilbert}, {Jenkner}, {Khandrika}, {Levay}, {Lucas}, {MacKenty},
  {Ogaz}, {Porterfield}, {Reid}, {Robberto}, {Royle}, {Smith},
  {Storrie-Lombardi}, {Sunnquist}, {Surace}, {Taylor}, {Williams}, {Bullock},
  {Dickinson}, {Finkelstein}, {Natarajan}, {Richard}, {Robertson}, {Tumlinson},
  {Zitrin}, {Flanagan}, {Sembach}, {Soifer}, \& {Mountain}}]{Lotz_2017}
{Lotz}, J.~M., {Koekemoer}, A., {Coe}, D., {et~al.} 2017,
  \href{http://dx.doi.org/10.3847/1538-4357/837/1/97}{\JournalTitle{\apj}, 837,
  97}

\bibitem[{{Miknaitis} {et~al.}(2007){Miknaitis}, {Pignata}, {Rest},
  {Wood-Vasey}, {Blondin}, {Challis}, {Smith}, {Stubbs}, {Suntzeff}, {Foley},
  {Matheson}, {Tonry}, {Aguilera}, {Blackman}, {Becker}, {Clocchiatti},
  {Covarrubias}, {Davis}, {Filippenko}, {Garg}, {Garnavich}, {Hicken}, {Jha},
  {Krisciunas}, {Kirshner}, {Leibundgut}, {Li}, {Miceli}, {Narayan}, {Prieto},
  {Riess}, {Salvo}, {Schmidt}, {Sollerman}, {Spyromilio}, \&
  {Zenteno}}]{Miknaitis_2007}
{Miknaitis}, G., {Pignata}, G., {Rest}, A., {et~al.} 2007,
  \href{http://dx.doi.org/10.1086/519986}{\JournalTitle{\apj}, 666, 674}

\bibitem[{{Noeske} {et~al.}(2007){Noeske}, {Faber}, {Weiner}, {Koo}, {Primack},
  {Dekel}, {Papovich}, {Conselice}, {Le Floc'h}, {Rieke}, {Coil}, {Lotz},
  {Somerville}, \& {Bundy}}]{Noeske_2007}
{Noeske}, K.~G., {Faber}, S.~M., {Weiner}, B.~J., {et~al.} 2007,
  \href{http://dx.doi.org/10.1086/517927}{\JournalTitle{\apjl}, 660, L47}

\bibitem[{{Peng} {et~al.}(2002){Peng}, {Ho}, {Impey}, \& {Rix}}]{Peng_2002}
{Peng}, C.~Y., {Ho}, L.~C., {Impey}, C.~D., \& {Rix}, H.-W. 2002,
  \href{http://dx.doi.org/10.1086/340952}{\JournalTitle{\aj}, 124, 266}

\bibitem[{{Peng} {et~al.}(2010){Peng}, {Ho}, {Impey}, \& {Rix}}]{Peng_2010}
---. 2010,
  \href{http://dx.doi.org/10.1088/0004-6256/139/6/2097}{\JournalTitle{\aj},
  139, 2097}

\bibitem[{{Persson} {et~al.}(2008){Persson}, {Barkhouser}, {Birk}, {Hammond},
  {Harding}, {Koch}, {Marshall}, {McCarthy}, {Murphy}, {Orndorff},
  {Scharfstein}, {Shectman}, {Smee}, \& {Uomoto}}]{Persson_2008}
{Persson}, S.~E., {Barkhouser}, R., {Birk}, C., {et~al.} 2008,
  \href{http://dx.doi.org/10.1117/12.790015}{in Society of Photo-Optical
  Instrumentation Engineers (SPIE) Conference Series, Vol. 7014, Ground-based
  and Airborne Instrumentation for Astronomy II}, 70142V

\bibitem[{{Petrillo} {et~al.}(2019){Petrillo}, {Tortora}, {Vernardos},
  {Koopmans}, {Verdoes Kleijn}, {Bilicki}, {Napolitano}, {Chatterjee},
  {Covone}, {Dvornik}, {Erben}, {Getman}, {Giblin}, {Heymans}, {de Jong},
  {Kuijken}, {Schneider}, {Shan}, {Spiniello}, \& {Wright}}]{Petrillo_2019}
{Petrillo}, C.~E., {Tortora}, C., {Vernardos}, G., {et~al.} 2019,
  \href{http://dx.doi.org/10.1093/mnras/stz189}{\JournalTitle{\mnras}, 484,
  3879}

\bibitem[{{Pettini} {et~al.}(2000){Pettini}, {Steidel}, {Adelberger},
  {Dickinson}, \& {Giavalisco}}]{Pettini_2000}
{Pettini}, M., {Steidel}, C.~C., {Adelberger}, K.~L., {Dickinson}, M., \&
  {Giavalisco}, M. 2000,
  \href{http://dx.doi.org/10.1086/308176}{\JournalTitle{\apj}, 528, 96}

\bibitem[{{Prescod-Weinstein} {et~al.}(2020){Prescod-Weinstein}, {Walkowicz},
  {Tuttle}, {Nord}, \& {Neilson}}]{2020arXiv200100674P}
{Prescod-Weinstein}, C., {Walkowicz}, L.~M., {Tuttle}, S., {Nord}, B., \&
  {Neilson}, H.~R. 2020, \JournalTitle{arXiv e-prints}, arXiv:2001.00674

\bibitem[{{Rest} {et~al.}(2005){Rest}, {Stubbs}, {Becker}, {Miknaitis},
  {Miceli}, {Covarrubias}, {Hawley}, {Smith}, {Suntzeff}, {Olsen}, {Prieto},
  {Hiriart}, {Welch}, {Cook}, {Nikolaev}, {Huber}, {Prochtor}, {Clocchiatti},
  {Minniti}, {Garg}, {Challis}, {Keller}, \& {Schmidt}}]{Rest_2005}
{Rest}, A., {Stubbs}, C., {Becker}, A.~C., {et~al.} 2005,
  \href{http://dx.doi.org/10.1086/497060}{\JournalTitle{\apj}, 634, 1103}

\bibitem[{{Richard} {et~al.}(2010){Richard}, {Smith}, {Kneib}, {Ellis},
  {Sanderson}, {Pei}, {Targett}, {Sand}, {Swinbank}, {Dannerbauer}, {Mazzotta},
  {Limousin}, {Egami}, {Jullo}, {Hamilton-Morris}, \& {Moran}}]{Richard_2010}
{Richard}, J., {Smith}, G.~P., {Kneib}, J.-P., {et~al.} 2010,
  \href{http://dx.doi.org/10.1111/j.1365-2966.2009.16274.x}{\JournalTitle{\mnras},
  404, 325}

\bibitem[{{Riechers} {et~al.}(2013){Riechers}, {Bradford}, {Clements},
  {Dowell}, {P{\'e}rez-Fournon}, {Ivison}, {Bridge}, {Conley}, {Fu}, {Vieira},
  {Wardlow}, {Calanog}, {Cooray}, {Hurley}, {Neri}, {Kamenetzky}, {Aguirre},
  {Altieri}, {Arumugam}, {Benford}, {B{\'e}thermin}, {Bock}, {Burgarella},
  {Cabrera-Lavers}, {Chapman}, {Cox}, {Dunlop}, {Earle}, {Farrah}, {Ferrero},
  {Franceschini}, {Gavazzi}, {Glenn}, {Solares}, {Gurwell}, {Halpern},
  {Hatziminaoglou}, {Hyde}, {Ibar}, {Kov{\'a}cs}, {Krips}, {Lupu}, {Maloney},
  {Martinez-Navajas}, {Matsuhara}, {Murphy}, {Naylor}, {Nguyen}, {Oliver},
  {Omont}, {Page}, {Petitpas}, {Rangwala}, {Roseboom}, {Scott}, {Smith},
  {Staguhn}, {Streblyanska}, {Thomson}, {Valtchanov}, {Viero}, {Wang},
  {Zemcov}, \& {Zmuidzinas}}]{Riechers_2013}
{Riechers}, D.~A., {Bradford}, C.~M., {Clements}, D.~L., {et~al.} 2013,
  \href{http://dx.doi.org/10.1038/nature12050}{\JournalTitle{\nat}, 496, 329}

\bibitem[{{Rigby} {et~al.}(2018{\natexlab{a}}){Rigby}, {Bayliss}, {Sharon},
  {Gladders}, {Chisholm}, {Dahle}, {Johnson}, {Paterno-Mahler}, {Wuyts}, \&
  {Kelson}}]{Rigby_2018a}
{Rigby}, J.~R., {Bayliss}, M.~B., {Sharon}, K., {et~al.} 2018{\natexlab{a}},
  \href{http://dx.doi.org/10.3847/1538-3881/aaa2ff}{\JournalTitle{\aj}, 155,
  104}

\bibitem[{{Rigby} {et~al.}(2018{\natexlab{b}}){Rigby}, {Bayliss}, {Chisholm},
  {Bordoloi}, {Sharon}, {Gladders}, {Johnson}, {Paterno-Mahler}, {Wuyts},
  {Dahle}, \& {Acharyya}}]{Rigby_2018b}
{Rigby}, J.~R., {Bayliss}, M.~B., {Chisholm}, J., {et~al.} 2018{\natexlab{b}},
  \href{http://dx.doi.org/10.3847/1538-4357/aaa2fc}{\JournalTitle{\apj}, 853,
  87}

\bibitem[{{Rivera-Thorsen} {et~al.}(2019){Rivera-Thorsen}, {Dahle}, {Chisholm},
  {Florian}, {Gronke}, {Rigby}, {Gladders}, {Mahler}, {Sharon}, \&
  {Bayliss}}]{Rivera-Thorsen_2019}
{Rivera-Thorsen}, T.~E., {Dahle}, H., {Chisholm}, J., {et~al.} 2019,
  \href{http://dx.doi.org/10.1126/science.aaw0978}{\JournalTitle{Science}, 366,
  738}

\bibitem[{{Rodighiero} {et~al.}(2011){Rodighiero}, {Daddi}, {Baronchelli},
  {Cimatti}, {Renzini}, {Aussel}, {Popesso}, {Lutz}, {Andreani}, {Berta},
  {Cava}, {Elbaz}, {Feltre}, {Fontana}, {F{\"o}rster Schreiber},
  {Franceschini}, {Genzel}, {Grazian}, {Gruppioni}, {Ilbert}, {Le Floch},
  {Magdis}, {Magliocchetti}, {Magnelli}, {Maiolino}, {McCracken}, {Nordon},
  {Poglitsch}, {Santini}, {Pozzi}, {Riguccini}, {Tacconi}, {Wuyts}, \&
  {Zamorani}}]{Rodighiero_2011}
{Rodighiero}, G., {Daddi}, E., {Baronchelli}, I., {et~al.} 2011,
  \href{http://dx.doi.org/10.1088/2041-8205/739/2/L40}{\JournalTitle{\apjl},
  739, L40}

\bibitem[{{Rosa-Gonz{\'a}lez} {et~al.}(2002){Rosa-Gonz{\'a}lez}, {Terlevich},
  \& {Terlevich}}]{2002MNRAS.332..283R}
{Rosa-Gonz{\'a}lez}, D., {Terlevich}, E., \& {Terlevich}, R. 2002,
  \href{http://dx.doi.org/10.1046/j.1365-8711.2002.05285.x}{\JournalTitle{\mnras},
  332, 283}

\bibitem[{{Salmon} {et~al.}(2020){Salmon}, {Coe}, {Bradley}, {Bouwens},
  {Brada{\v{c}}}, {Huang}, {Oesch}, {Stark}, {Sharon}, {Trenti}, {Avila},
  {Ogaz}, {Andrade-Santos}, {Carrasco}, {Cerny}, {Dawson}, {Frye}, {Hoag},
  {Johnson}, {Jones}, {Lam}, {Lovisari}, {Mainali}, {Past}, {Paterno-Mahler},
  {Peterson}, {Riess}, {Rodney}, {Ryan}, {Sendra-Server}, {Strait}, {Strolger},
  {Umetsu}, {Vulcani}, \& {Zitrin}}]{Salmon_2020}
{Salmon}, B., {Coe}, D., {Bradley}, L., {et~al.} 2020,
  \href{http://dx.doi.org/10.3847/1538-4357/ab5a8b}{\JournalTitle{\apj}, 889,
  189}

\bibitem[{{Santini} {et~al.}(2017){Santini}, {Fontana}, {Castellano}, {Di
  Criscienzo}, {Merlin}, {Amorin}, {Cullen}, {Daddi}, {Dickinson}, {Dunlop},
  {Grazian}, {Lamastra}, {McLure}, {Micha{\l}owski}, {Pentericci}, \&
  {Shu}}]{Santini_2017}
{Santini}, P., {Fontana}, A., {Castellano}, M., {et~al.} 2017,
  \href{http://dx.doi.org/10.3847/1538-4357/aa8874}{\JournalTitle{\apj}, 847,
  76}

\bibitem[{{Schechter}(1976)}]{1976ApJ...203..297S}
{Schechter}, P. 1976,
  \href{http://dx.doi.org/10.1086/154079}{\JournalTitle{\apj}, 203, 297}

\bibitem[{{Shapley} {et~al.}(2003){Shapley}, {Steidel}, {Pettini}, \&
  {Adelberger}}]{Shapley_2003}
{Shapley}, A.~E., {Steidel}, C.~C., {Pettini}, M., \& {Adelberger}, K.~L. 2003,
  \href{http://dx.doi.org/10.1086/373922}{\JournalTitle{\apj}, 588, 65}

\bibitem[{{Sharon} {et~al.}(2020){Sharon}, {Bayliss}, {Dahle}, {Dunham},
  {Florian}, {Gladders}, {Johnson}, {Mahler}, {Paterno-Mahler}, {Rigby},
  {Whitaker}, {Akhshik}, {Koester}, {Murray}, {Remolina Gonzalez}, \&
  {Wuyts}}]{Sharon_2020}
{Sharon}, K., {Bayliss}, M.~B., {Dahle}, H., {et~al.} 2020,
  \JournalTitle{VizieR Online Data Catalog}, J/ApJS/247/12

\bibitem[{{Soifer} {et~al.}(1998){Soifer}, {Neugebauer}, {Franx}, {Matthews},
  \& {Illingworth}}]{Soifer_1998}
{Soifer}, B.~T., {Neugebauer}, G., {Franx}, M., {Matthews}, K., \&
  {Illingworth}, G.~D. 1998,
  \href{http://dx.doi.org/10.1086/311462}{\JournalTitle{\apjl}, 501, L171}

\bibitem[{{Stalder} {et~al.}(2014){Stalder}, {Stark}, {Amato}, {Geary},
  {Shectman}, {Stubbs}, \& {Szentgyorgyi}}]{Stalder_2014}
{Stalder}, B., {Stark}, A.~A., {Amato}, S.~M., {et~al.} 2014,
  \href{http://dx.doi.org/10.1117/12.2054933}{in Society of Photo-Optical
  Instrumentation Engineers (SPIE) Conference Series, Vol. 9147, Ground-based
  and Airborne Instrumentation for Astronomy V}, 91473Y

\bibitem[{{Stark} {et~al.}(2013){Stark}, {Auger}, {Belokurov}, {Jones},
  {Robertson}, {Ellis}, {Sand }, {Moiseev}, {Eagle}, \& {Myers}}]{Stark_2013}
{Stark}, D.~P., {Auger}, M., {Belokurov}, V., {et~al.} 2013,
  \href{http://dx.doi.org/10.1093/mnras/stt1624}{\JournalTitle{\mnras}, 436,
  1040}

\bibitem[{{Steidel} {et~al.}(2016){Steidel}, {Strom}, {Pettini}, {Rudie},
  {Reddy}, \& {Trainor}}]{Steidel_2016}
{Steidel}, C.~C., {Strom}, A.~L., {Pettini}, M., {et~al.} 2016,
  \href{http://dx.doi.org/10.3847/0004-637X/826/2/159}{\JournalTitle{\apj},
  826, 159}

\bibitem[{{Strandet} {et~al.}(2017){Strandet}, {Weiss}, {De Breuck}, {Marrone},
  {Vieira}, {Aravena}, {Ashby}, {B{\'e}thermin}, {Bothwell}, {Bradford},
  {Carlstrom}, {Chapman}, {Cunningham}, {Chen}, {Fassnacht}, {Gonzalez},
  {Greve}, {Gullberg}, {Hayward}, {Hezaveh}, {Litke}, {Ma}, {Malkan}, {Menten},
  {Miller}, {Murphy}, {Narayanan}, {Phadke}, {Rotermund}, {Spilker}, \&
  {Sreevani}}]{Strandet_2017}
{Strandet}, M.~L., {Weiss}, A., {De Breuck}, C., {et~al.} 2017,
  \href{http://dx.doi.org/10.3847/2041-8213/aa74b0}{\JournalTitle{\apjl}, 842,
  L15}

\bibitem[{{Verdugo} {et~al.}(2011){Verdugo}, {Motta}, {Mu{\~n}oz}, {Limousin},
  {Cabanac}, \& {Richard}}]{Verdugo_2011}
{Verdugo}, T., {Motta}, V., {Mu{\~n}oz}, R.~P., {et~al.} 2011,
  \href{http://dx.doi.org/10.1051/0004-6361/201014965}{\JournalTitle{\aap},
  527, A124}

\end{thebibliography}

\pagebreak
\pagebreak

\end{document}